\documentclass[rmp,aps,preprint,nofootinbib,endfloats]{revtex4}

\usepackage{amsmath}
\usepackage{epsfig}
\usepackage{rotating}

\newcommand*{\ioo}{$\langle 100 \rangle$}
\newcommand*{\iio}{$\langle 110 \rangle$}
\newcommand*{\iii}{$\langle 111 \rangle$}
\newcommand*{\iiz}{$\langle 112 \rangle$}

\begin{document}

\title{{\em Colloquium}: Structural, electronic and 
       transport properties of silicon nanowires}

\author{Riccardo Rurali}
\email{rrurali@icmab.es}
\affiliation{Departament d'Enginyeria Electr\`{o}nica,
             Universitat Aut\`{o}noma de Barcelona,
             08193 Bellaterra, Spain}
\affiliation{Institut de Ci\`{e}ncia de Materials
             de Barcelona (CSIC), Campus de Bellaterra,
             08193 Bellaterra, Barcelona, Spain}

\date{\today}

\begin{abstract}
In this paper we review the theory of silicon nanowires. 
We focus on nanowires with diameters below 10~nm, where quantum 
effects become important and the properties diverge 
significantly from those of bulk silicon. These wires can
be efficiently treated within electronic structure 
simulation methods and will be among the most important
functional blocks of future nanoelectronic devices. 
Firstly, we review the structural properties of silicon
nanowires, emphasizing the close connection between
the growth orientation, the cross-section and the 
bounding facets. Secondly, we discuss the electronic
structure of pristine and doped nanowires, which hold
the ultimate key for their applicability in novel 
electronic devices. Finally, we review transport properties 
where some of the most important limitations in the
performances of nanowire-based devices can lay.
Many of the unique properties of these systems are
at the same time defying challenges and opportunities 
for great technological advances. 
\end{abstract}


\maketitle

\tableofcontents

\section{Introduction}

One-dimensional nanostructured systems have attracted a 
great attention in the last two decades, with this 
interest extraordinary boosted by the facile synthesis of 
carbon nanotubes (CNTs) reported in the beginning 
of the 90s~\cite{IijimaNature91}. 
The reason is twofold: on the one hand they have proved to 
be an excellent test-bed to study the most intriguing physical 
effects, whereas on the other hand they are believed
to be among the most important building blocks of the
next generation of electronic devices.

CNTs are hollow cylinders obtained by rolling up one 
or more graphene sheets, a one-atom-thick allotrope of 
carbon~\cite{CharlierRMP08}. The symmetry and the 
electronic structure of graphene~\cite{CastroNetoRMP09} 
are such that the properties of the CNT depends critically
on the {\em exact} way it is wrapped up,
and it can be either metallic or semiconducting. 
This confer the CNTs with a richer physics, but it is
clearly far from ideal from the viewpoint of applications,
especially when --as it is the case-- a simple route to 
selectively grow one type of CNT or the other is lacking.

Nanowires are an extremely attractive alternative to CNTs,
because it is much easier to control their electrical
properties and, as long as the surface is properly passivated 
--something that occurs naturally during or right after 
growth--, they are invariably semiconducting~\footnote{Below 
we will discuss explicitly a few cases where 
nanowires derived from semiconducting solids can be
metallic, whereas nanowires made of metal atoms are beyond
the scope of this paper and are not being discussed.}.

Silicon nanowires 
(SiNWs), in particular, look like a very appealing choice, 
since they provide the ideal interface with the 
existing Si devices, while taking advantage from a 
tractable material technology. SiNWs are commonly grown
by the vapor-liquid-solid technique~\cite{WagnerAPL64,
WestwaterJVSTB97}, where a Au nanoparticle is used to
catalyze SiH$_4$ decomposition. Briefly, the Au particle
is deposited onto a Si surface and react with the Si
atoms of the substrate, forming Au-Si alloy droplets.
These droplets adsorb Si from the vapor phase, resulting
in a supersaturated state where the Si atoms precipitate
and the SiNW starts nucleating~\footnote{See \textcite{WangMatSciEng08}
for a comprehensive review of the growth techniques.}.
 
\begin{figure}
\centering
\epsfig{file=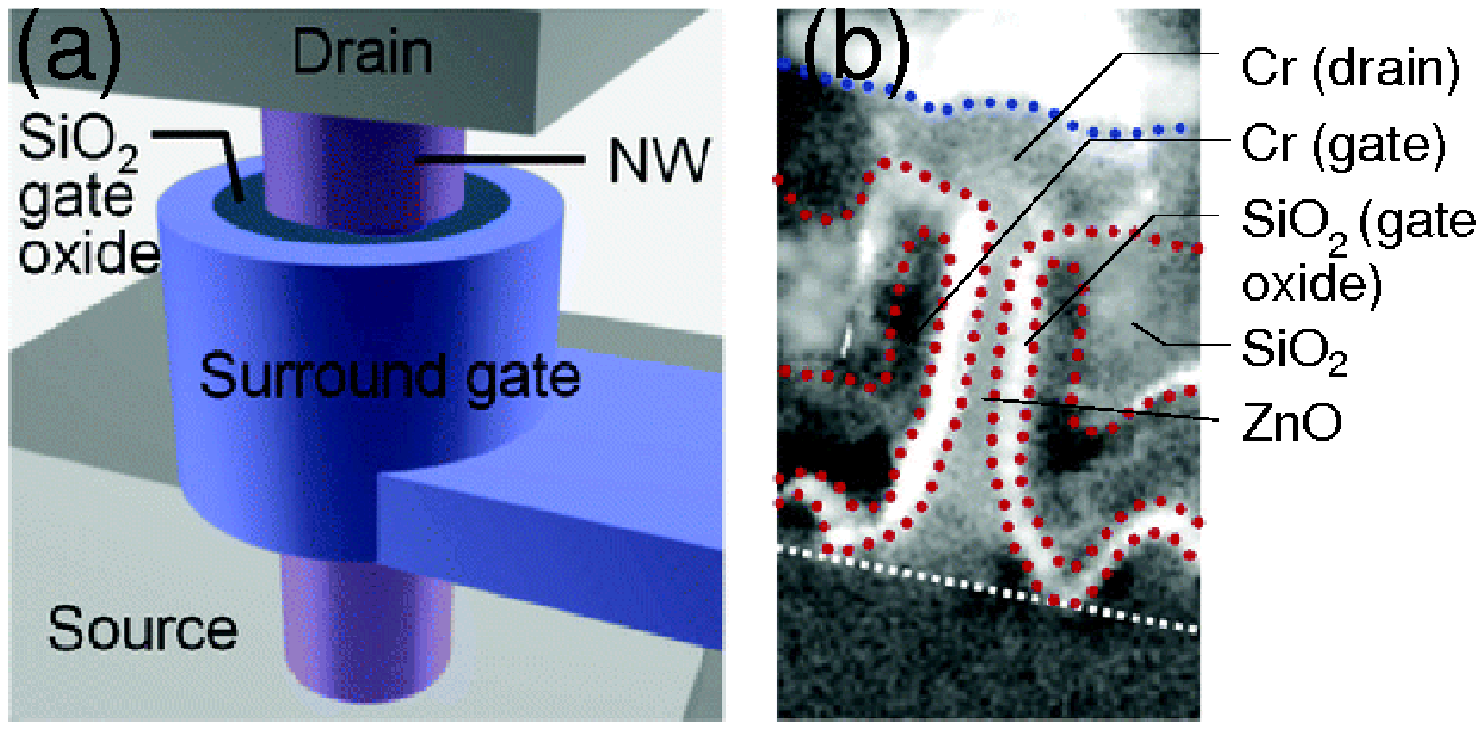, width=1.00\linewidth, clip=, angle=0}
\vskip 0cm
\caption{(Color online) (a)~Cartoon and (b)~experimental 
         realization of a ZnO nanowire-based field-effect 
         transistor with an {\em all-around} (or surrounding) 
         gate. The channel length is 200 nm.
         From \textcite{NgNL04}. \vskip 5cm}
\label{fig:NgNL04fig0}
\end{figure}

As David K. Ferry illustrates in an enlightening
paper~\cite{FerryScience08}, nanowires could provide 
the paradigm shift needed to continue improving the 
density and the performances of electronic circuits.
For almost four decades the
increase in computing power has been described by the
well-known Moore's law~\cite{MooreEl65}, which has 
been standing on three pillars: (a)~the 
increase of the size of the microchips;
(b)~the reduction of the transistor size, and (c)~the
{\em circuit cleverness}, that is the reduction of the
number of devices required to perform a certain function.
While the first of these driving forces played a
significant role only in the pioneering years of
solid state electronics, the reduction of device size
has a pivotal role, since the physical limit of 
material scaling is rapidly approaching.
Nanowires can lead to an obvious benefit concerning
the miniaturization, thanks to bottom-up growth
that allows overcoming the limit of conventional
lithography-based top-down design. 
Subtler are the perspective advantages concerning circuit 
cleverness, which can be significantly improved by taking 
advantage of the coexisting nature of interconnection 
and active device of nanowires. 
In particular,
a replacement of metallic {\em vias} with vertical
transistors is envisaged. These
new circuits could be easily
reconfigurated to perform different operations, achieving
a much higher level of integration~\cite{FerryScience08}.
Additionally, compared to classical planar device technology, 
nanowires can better accommodate {\em all-around} gates (see 
Fig.~\ref{fig:NgNL04fig0}), 
which improve field-effect efficiency and device 
performances~\cite{ColingeSSE04,NgNL04}
and mobilities of $\sim$~1000~cm$^2$V$^{-1}$s$^{-1}$, 
substantially larger than those obtained in conventional 
Si devices, have been obtained~\cite{CuiNL03}~\footnote{It 
is difficult to make rigorous comparisons, because the
mobility has a strong inverse dependence on the dopant
density which is seldom known with accuracy in nanowires.
However, the peak value of 1350~cm$^2$V$^{-1}$s$^{-1}$ 
obtained for the hole mobility by \textcite{CuiNL03} must be 
compared with the typical values for bulk Si of 
$\sim$~400~cm$^2$V$^{-1}$s$^{-1}$ and 
$\sim$~100~cm$^2$V$^{-1}$s$^{-1}$ for an acceptor 
concentration of 10$^{16}$~cm$^{-3}$ and 10$^{18}$~cm$^{-3}$, 
respectively.}.

Several promising applications have already been demonstrated,
ranging from electron devices~\cite{ChungAPL00,YuJPCB00,
CuiScience01,CuiNL03,ZhengAdvMat04,GoldbergerNL06,HuNL08,
WuIEEE08,WangNL06b}, logic gates~\cite{HuangScience01}, non-volatile
memories~\cite{DuanNL02}, photovoltaics~\cite{TianNature07,
KempaNL08,TianCSR08}, photonics~\cite{GudiksenNature02,
PauzauskieMatTod06}, to biological sensors~\cite{CuiScience01b,
ZhongScience03,HahmNL04}.
On top of that, giant piezoresistance effect~\cite{HeNatNano06}
and enhanced thermoelectric performances~\cite{HochbaumNature08,
BoukaiNature08} have recently been reported.
The interested reader is encouraged to check some of the 
many experimental reviews~\cite{PatolskyMatTod05,
LiMatTod06,ThelanderMatTod06,LuJPD06,
WuCM08,XiaAdvMat03,Nanosilicon}.

In this paper we will review the theory of SiNWs. 
Clearly, we will make several references to experiments, 
whenever they support or challenge the theoretical predictions.
Sometimes the comparisons 
are difficult to make, because SiNWs that are routinely 
grown range from 50 to 200~nm, while those 
that can be efficiently studied within electronic 
structure methods are 2-3~nm thick, at most.
Luckily, this gap is slowly narrowing and thin SiNWs
with diameters below 10~nm have been successfully
grown by several groups~\cite{MoralesScience98,HolmesScience00,
CuiAPL01,ColemanJACS01,ColemanJACS01b,MaScience03,CuiNL03,
WuNL04,ZhongNL05,DePadovaNL08}. The theoretical results that we 
discuss outline the most urgent problems that will have 
to be dealt with within the next generation of nanowires, 
those with characteristic sizes approaching the quantum 
limit.

Although many of the features that we will discuss are 
common to other types of semiconducting nanowires, for 
the sake of clarity we will restrict to SiNWs throughout 
the paper. It should be at least pointed out, however, 
that in recent years tremendous progresses are being made 
with compound semiconductors nanowires --mainly III-V 
nanowires-- especially for what concerns photonics 
application~\cite{BjorkAPL02,ThelanderAPL03,DickNatMat04}.

A final remark concerns the computational methodologies. 
Although the main goal of the paper is giving a complete 
overview of the most important results that have been 
obtained within atomistic simulations, we will not enter 
in technical details, unless where it is necessary. Most 
of the results have
been obtained within density functional theory (DFT), whose
theoretical grounds are clearly out of the scope of this work.
The interested reader can look at both the original 
papers~\cite{HohenbergPR64,KohnPR65}, excellent 
reviews~\cite{JonesRMP89,PayneRMP92} and comprehensive 
books~\cite{Martin}. Less frequently, we will refer to the
tight-binding formalism~\cite{SlaterPR54,GoringeRPP97,
ColomboRNC05} or to empirical interatomic 
potentials~\cite{TersoffPRB89,StillingerPRB85,JustoPRB98}.

\section{Structural properties}
\label{sec:struct}

\subsection{Growth orientations and monocrystallinity}
\label{sec:axis}

The extraordinary impact that the discovery of carbon
nanotubes~\cite{RadushkevichZFC52,OberlinJCG76,IijimaNature91} 
had on condensed matter and nanoscience at first biased the 
research on Si quasi one-dimensional systems to the 
pursuit of tubular structures. Hollow structures
resembling carbon nanotubes~\cite{LiPRB02}, structures 
based on hollow elements~\cite{MenonPRL99} or on
fullerene-like system~\cite{MarsenPRB99} have been 
proposed. Although these --or other structures inspired 
by cluster assemble~\cite{SenPRB02}-- are stable within a 
total energy framework, they have not been observed
experimentally to date.

In the meanwhile Si nanotubes have been indeed successfully 
synthesized~\cite{ShaAdvMat02}, while their use for 
nanoelectronics still remains troublesome~\cite{PerepichkaSmall06}, 
and things with Si nanowires turned out to be simpler than 
speculated. Convincing experimental evidence soon indicated that 
SiNWs are rod-like structures constructed around a bulk Si 
single-crystalline core~\cite{MoralesScience98,ZhangPRB00,
HolmesScience00,TeoNL03,WuNL04}.

\begin{figure}
\centering
\epsfig{file=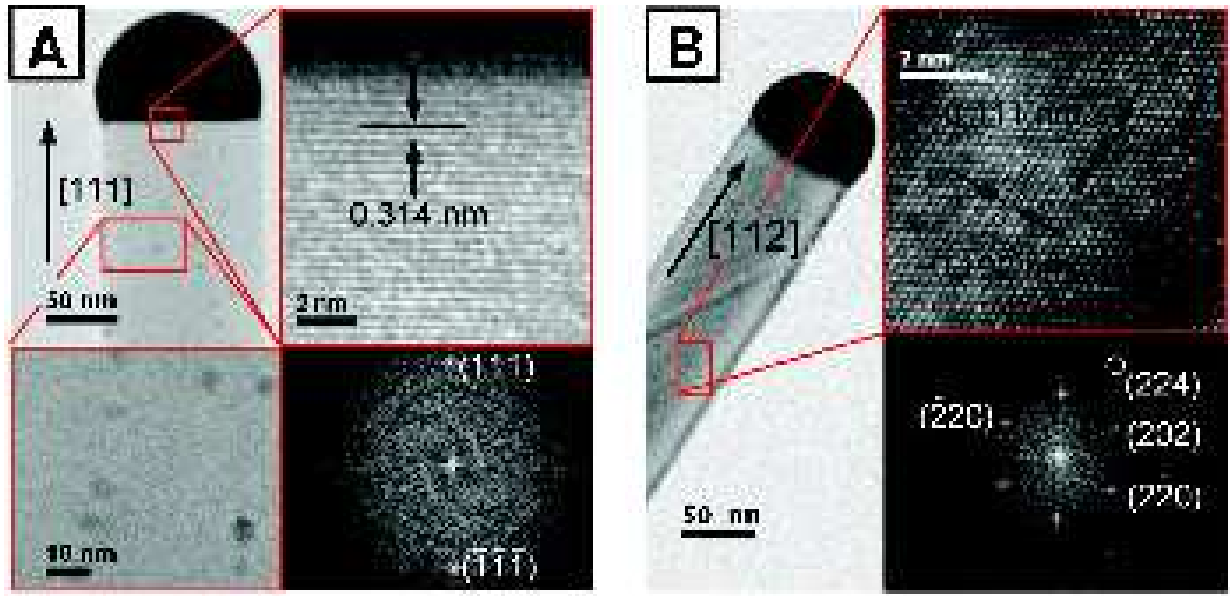, width=1.00\linewidth, clip=, angle=0}
\vskip 0cm
\caption{(Color online) Transmission electron microscopy
         (TEM) image of a single crystalline SiNW grown
         along (a)~the \iii\ and (b)~the \iiz\ axis. The
         high resolution TEM micrograph of the
         crystalline core shows clearly the Si(111) and
         Si(224) planes, respectively, together with the
         Fourier transform of the image. In case of the
         \iii\ SiNW a magnified view of the sidewalls of
         the wire show the presence of Au-Si particles
         about 7 nm in size. From \textcite{LugsteinNL08}.}
\label{fig:LugsteinNL08fig1c-2c}
\end{figure}

An important consequence of their single-crystal nature
is that SiNWs grow along very well defined crystalline 
directions (see Fig.~\ref{fig:LugsteinNL08fig1c-2c}).
\textcite{WuNL04} carried out an interesting and 
extensive study of the growth orientations,
showing a connection between the diameter and the favored 
crystal axis in the Au-catalyzed synthesis of SiNWs:
the smallest-diameter nanowires grow primarily along the 
\iio\ direction, whereas larger nanowires favor the \iii\ 
direction; intermediate diameters, 10 to 20~nm, on the other 
hand, are dominated by \iiz\ wires. 
Thermodynamic models have been proposed to account for this
diameter-dependent growth direction~\cite{SchmidtNL05,WangNL06} 
with consistent results in good agreement with the experiments, 
fixing the cross-over from \iio\ to \iii\ growth at 
20-25~nm (\iiz\ orientation was not considered in those studies).
The stacking sequence preference that leads to \iio\ over
\iii\ SiNWs at small diameters is also supported by 
first-principles calculations~\cite{AkiyamaPRB06}.
More recently, a continuum model that allows studying how 
growth begins and evolves toward steady-state wire growth 
has been presented~\cite{SchwarzPRL09}. The advantage of 
this approach is that complex situations such as catalyst 
coarsening and interrupted growth can be easily handled.

Ideally, nonetheless, one would like to be able 
to control the wire orientation at growth time.
An important achievement in this sense was the demonstration that the 
growth orientation can also be controlled externally by 
adjusting the growth pressure~\cite{HolmesScience00,
LugsteinNL08}. Alternatively, the use of different techniques
can bias somehow the growth along certain crystal axis. 
For instance, the less common oxide-assisted growth method, 
generally yielding a broader diameter distribution~\cite{WangPRB98}, 
might favor different orientations for ultra-thin 
SiNWs~\cite{TeoNL03}. Significantly, the thinnest SiNW reported 
to date~\cite{MaScience03} was synthesized with this technique 
and was a \iiz\ wire (see Fig.~\ref{fig:MaScience03fig1-2}). 

\subsection{Surface reconstructions in pristine nanowires}
\label{sub:pristine}

The next major issue one has to face when studying the structure
of a SiNW is the shape of its cross-section which, as we shall
see briefly, is intimately related with the growth orientation.
Although one can pictorially imagine nanowires as cylindrical 
structures, clearly, when going down to the atomic-scale detail,
this is not the structural arrangement that they assume --or even
{\em can} assume. The analogous problem in solids and small 
particles~\cite{WangCPL84,ZhdanovPRL98} is 
elegantly solved by means of the Wulff construction or 
Wulff rule~\cite{MarksRPP94}, which relates the equilibrium
shape with the surface free energy of the facets involved.
Solving the energy minimization problem $\text{min} \sum s \gamma_s$,
where $s$ is the number of surface unit cells and $\gamma_s$
the corresponding energy, leads to the optimum shape.

\textcite{ZhaoPRL03} have reexamined the use of
Wulff construction within the determination of the equilibrium
cross-sections of SiNWs. They showed that the conventional
formulation of the Wulff criterion lacks of two important
aspects: (i)~in solids and smooth spherical particles
the energy of the edges between facets is neglected
compared to the surface contribution; (ii)~the bulk is
assumed already at its minimum and thus invariant. 
Hence, they propose the following generalization for 
the Wulff energy:
\begin{equation}
F = E_e + \sum_s s \gamma_s + E_b
\label{eq:wulff}
\end{equation}
where they include the energy of matching adjacent facets 
$E_e$, i.e. the energy of the edges, and the energy of the 
bulk $E_b$, releasing the constraint on the innermost part 
of the wire which can now change. 

\begin{figure}
\centering
\epsfig{file=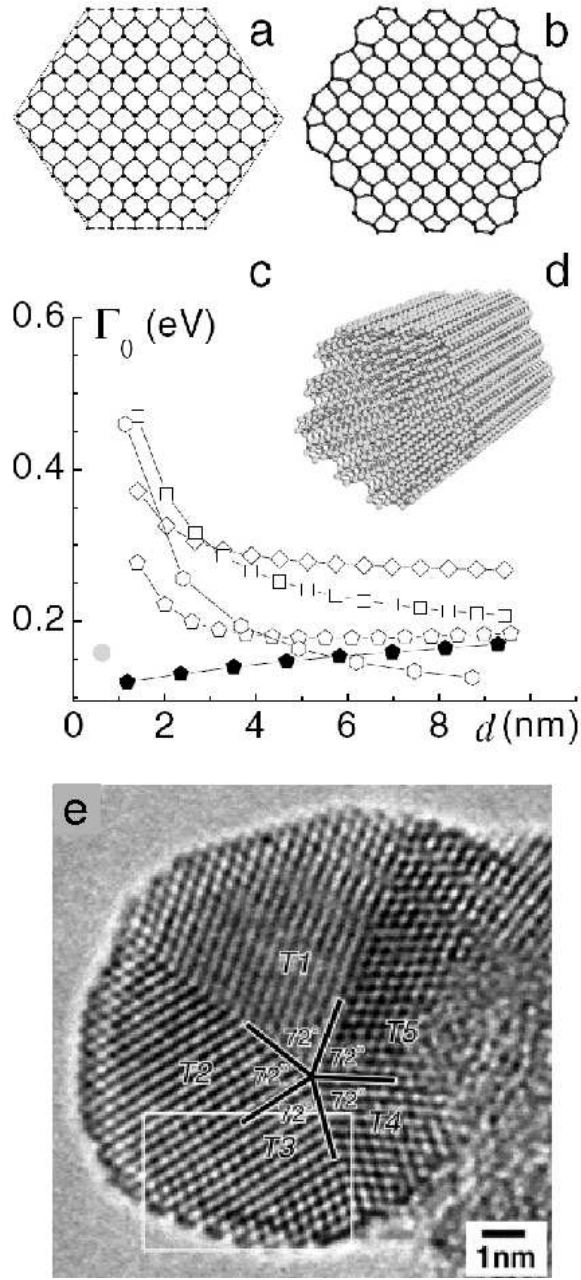, width=0.50\linewidth, clip=, angle=0}
\vskip 0cm
\caption{Hexagonal \iio\ wire with four \{111\} and two
         \{100\} (a)~unreconstructed and (b)~reconstructed
         facets. (c)~Energy of different types of wires
         (see \textcite{ZhaoPRL03}) as a function of their
         diameter $d$. The most stable structure for
         $d<6$~nm is [solid pentagons in panel (c)] is
         shown in panel (d). Panel (d) shows a high 
         resolution TEM image of a pentagonal nanowire 
         grown by \textcite{TakeguchiSurfSci01}.
         Adapted from \textcite{ZhaoPRL03}
         and \textcite{TakeguchiSurfSci01}.}
\label{fig:ZhaoPRL03fig2}
\end{figure}

\begin{figure}
\centering
\epsfig{file=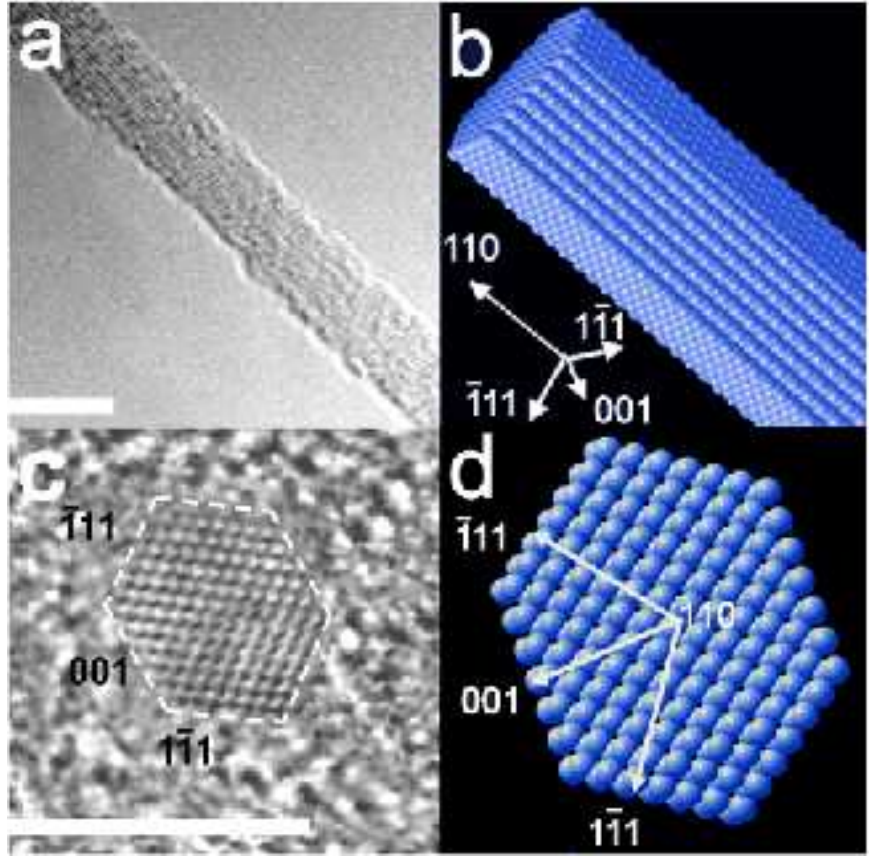, width=0.70\linewidth, clip=, angle=0}
\vskip 0cm
\caption{(Color online) (a)~TEM images of 3.8~nm SiNWs
          grown along the \iio\ direction, (c)~high resolution TEM
          cross-sectional image, and equilibrium shapes
          for the (b)~NW and the (d)~NW cross-sections
          predicted by Wulff construction. The scale bars
          are 5~nm. From \textcite{WuNL04}.}
\label{fig:WuNL04fig4}
\end{figure}

They investigated different faceting arrangements for 
SiNWs grown along the \iio\ axis comparing them 
on the basis of Eq.~\ref{eq:wulff}. They found that 
the ground-state structure for SiNWs up to 5~nm is a 
pentagonal cross-section constructed joining five prisms 
cut out of a [110] Si plane [see Fig.~\ref{fig:ZhaoPRL03fig2}(d)].
This structure has seldom been detected experimentally
[a remarkable observation by \textcite{TakeguchiSurfSci01} is 
shown in Fig.~\ref{fig:ZhaoPRL03fig2}(e)]),
probably because it is not constructed around a bulk-core,
which seems to be the favored situation at growth time. 
However, if one restricts to wires with a strictly bulk
core the model of \textcite{ZhaoPRL03} correctly predicts 
hexagonal over square cross-sections for \iio\ SiNWs, in 
agreement with the experiments~\cite{MaScience03,WuNL04}
(see Fig.~\ref{fig:WuNL04fig4}).

\begin{figure}
\centering
\epsfig{file=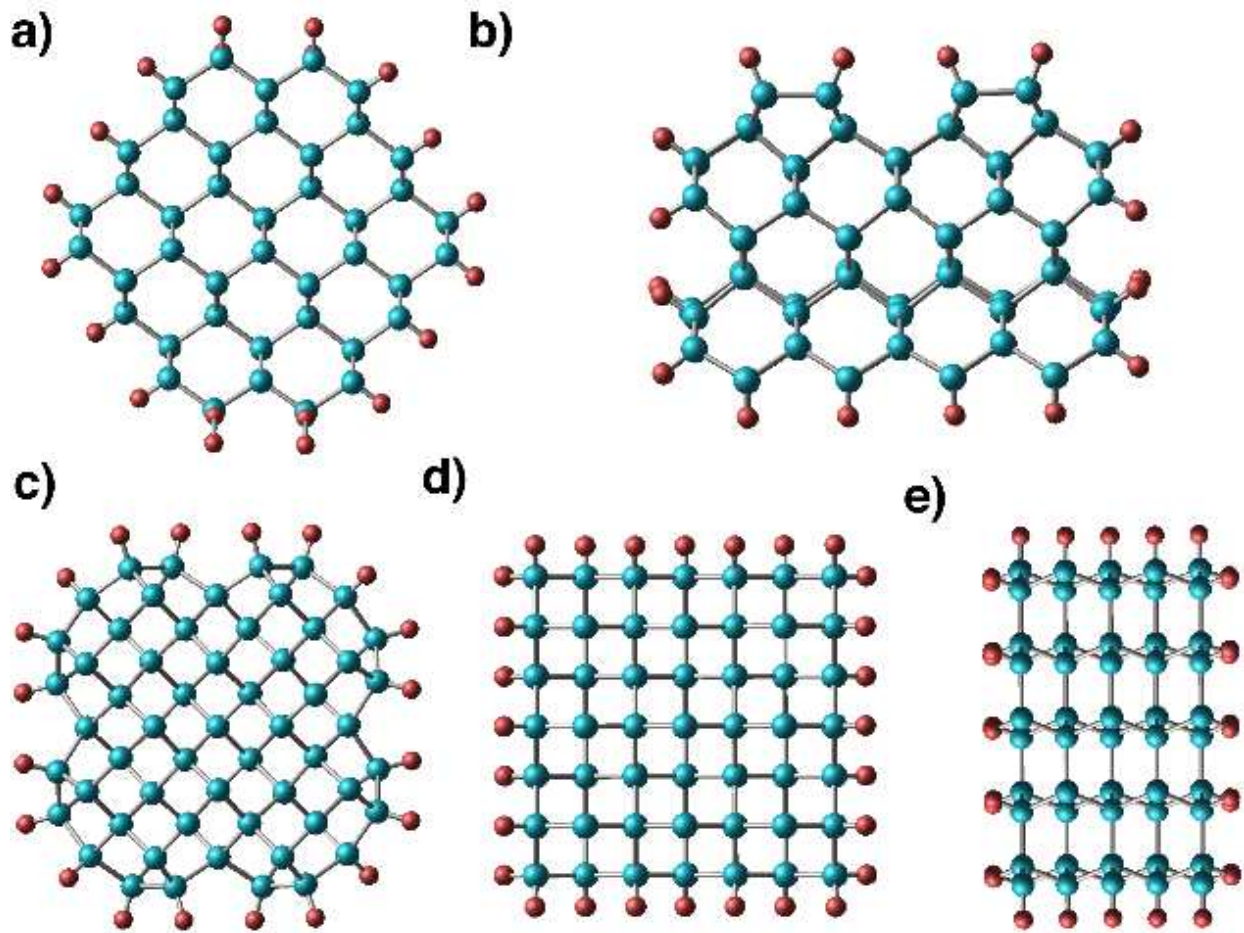, width=\linewidth, clip=, angle=0}
\vskip 0cm
\caption{(Color online) Optimized structures of possible
          cross-sections of H-passivated SiNWs grown along 
          (a,b) the \iio\, (c,d) the \ioo\ and (e) \iiz\ 
          orientation. Adapted from \textcite{SinghNL06}.}
\label{fig:SinghNL06fig1mod}
\end{figure}

\begin{figure}
\centering
\epsfig{file=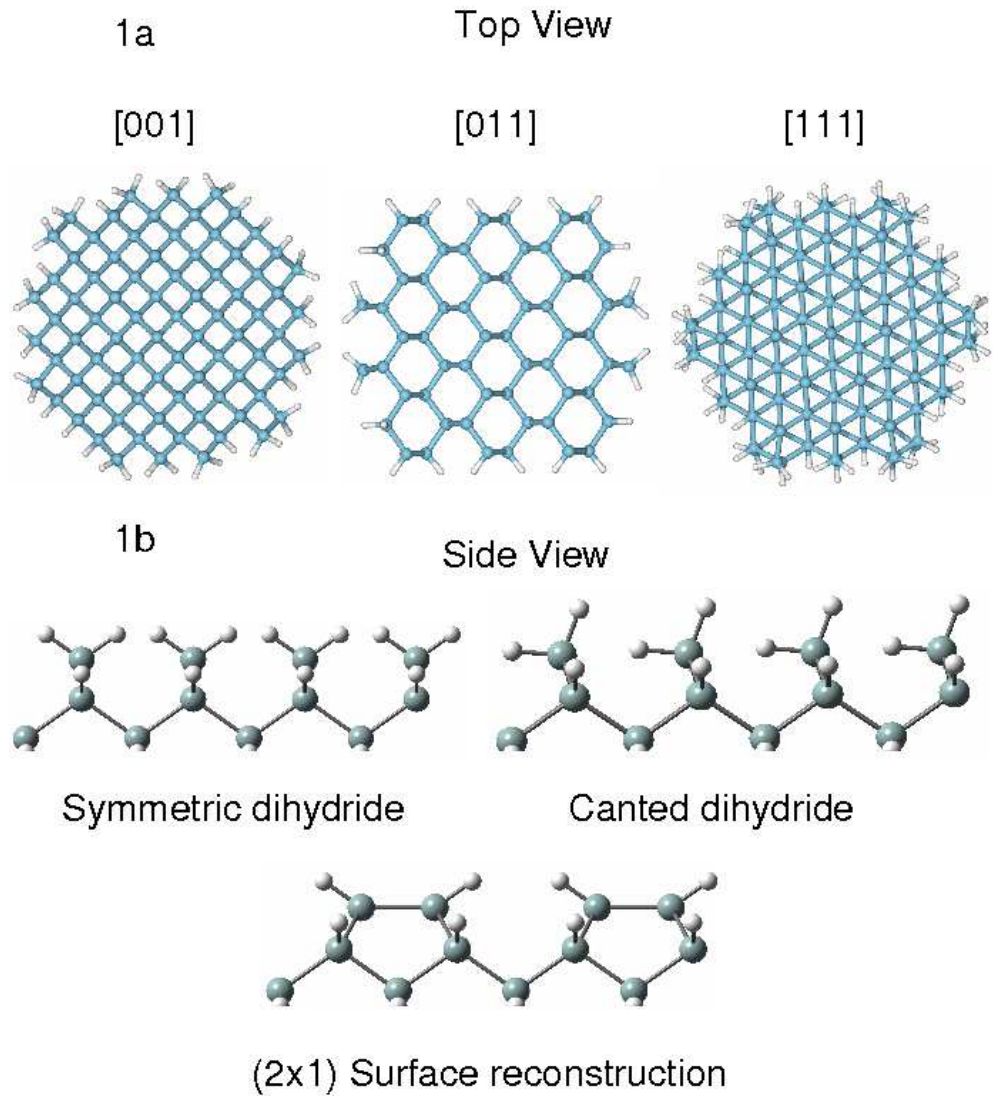, width=0.80\linewidth, clip=, angle=0}
\vskip 0cm
\caption{(Color online) (a)~Cross-section view of 3~nm SiNWs
          grown along three different directions \ioo, \iio,
          and \iii. (b)~Side view of three different surface
          structures; in the last configuration the surface
          first reconstructs and then is passivated.
          From \textcite{VoPRB06}.}
\label{fig:VoPRB06}
\end{figure}

The most important result of the work of \textcite{ZhaoPRL03}
is emphasizing the role of the edges and how the interplay
between edges and surfaces play a key role in determining
the reconstruction of Si one-dimensional structures.
Before its formalization, this effect had been already 
pointed out by \textcite{Ismail-BeigiPRB98} a few years
earlier. In their work they considered a pristine SiNW 
grown along the \ioo\ axis. This orientation favors 
a square cross-section with \{100\} facets, an energetically
{\em cheaper} solution than a square cross-section with \{111\} 
facets~\cite{RuraliNanotech05}. The abrupt match between 
the \{100\} facets results in an energetically expensive 
edge, a large value of $E_e$ in Eq.~\ref{eq:wulff}, which
can be reduced by forming smaller, transition \{110\} 
facets that allow a smoother match between the dominant 
\{100\} facets and partially release the stress 
accumulated at the edge (examples 
can be seen in Figs.~\ref{fig:SinghNL06fig1mod}(c),
\ref{fig:VoPRB06}(a) and \ref{fig:LeePRB07fig1-3}.).

Unfortunately, a word of care should be spent concerning
the above discussion. Down at the ultimate nanoscale
limit it is delicate to give general rules and for
extremely thin SiNWs counterexamples can be found to
the general trends discussed previously. For instance, 
\textcite{CaoPRL06} showed that the faceting arrangement 
proposed by \textcite{Ismail-BeigiPRB98} for \ioo\ wires 
and later followed by other authors~\cite{RuraliPRL05,
RuraliPRB05,VoPRB06,LeePRB07} is favored only beyond a 
1.7~nm diameter, whereas tiny SiNWs prefer sharp edges, 
i.e. the removal of the edges does not pay back.

In the spirit of the work of \textcite{ZhaoPRL04}, 
\textcite{JustoPRB07} carried out 
an interesting and systematic study of SiNWs grown
along the \ioo, \iio, and \iiz\ crystal axis,
carrying out extensive calculations based on 
an interatomic potential~\cite{JustoPRB98}. 
In order to elucidate the role of the different facets
for the stability, for each growth orientation they 
examined cross-sections bounded by different facet 
compositions. For the \ioo\ SiNWs, for instance, 
they considered both all-\{100\} facets, all-\{110\} 
facets and three intermediate combinations.
Proceeding in this way they were able to formulate
a universal scaling law in terms of the wire
perimeter, according to which the nanowire energy
per atom always lies within two limiting energy lines, 
which are directly related to the character of the 
prevailing facets. Interestingly, in the limit of thick wires,
the edge energy become negligible as suggested by 
\textcite{ZhaoPRL04} and the energy scales linearly
with the inverse of the wire perimeter.

Silicon has a very rich phase diagram~\cite{KaczmarskiPRL05}
and many solid phases other than the diamond structure
are known. Among them is the so-called clathrate phase 
that becomes stable at negative pressures.
The stability of such a phase for quasi one-dimensional
nanostructure has been investigated by Ponomareva and 
co-workers~\cite{PonomarevaPRL05,PonomarevaPRB06}. 
They studied cage-like SiNWs {\em carved out} 
of a Si clathrate structure and compared them with both 
tetrahedral diamond-like and polycrystalline SiNWs. 
Their results indicate that also in these nanostructures 
the tetrahedral structure is favored. Nevertheless, the 
difference in energy is rather small and it is suggested 
that clathrate based SiNWs might have better conductive 
properties.

\subsection{Passivated nanowires}
\label{sub:pass}

The study of the structure of pristine SiNWs has been a 
fertile ground where to start the theoretical research
of these fascinating systems. However, quite soon 
it became clear that the wires grown experimentally 
have always {\em passivated} facets. 
Silicon form highly directional covalent bonds
according to the know sp$^3$ tetrahedral pattern.
Silicon atoms at the surface have dangling bonds (DB),
unsaturated bonds that make the atom highly reactive
and that induce strong reconstruction of the surfaces.
Generally speaking, surface passivation consists of
the termination of DBs on the surface with 
elements that assure their chemical stability. Hence
the surface is chemically {\em passive}.

\begin{figure}
\centering
\epsfig{file=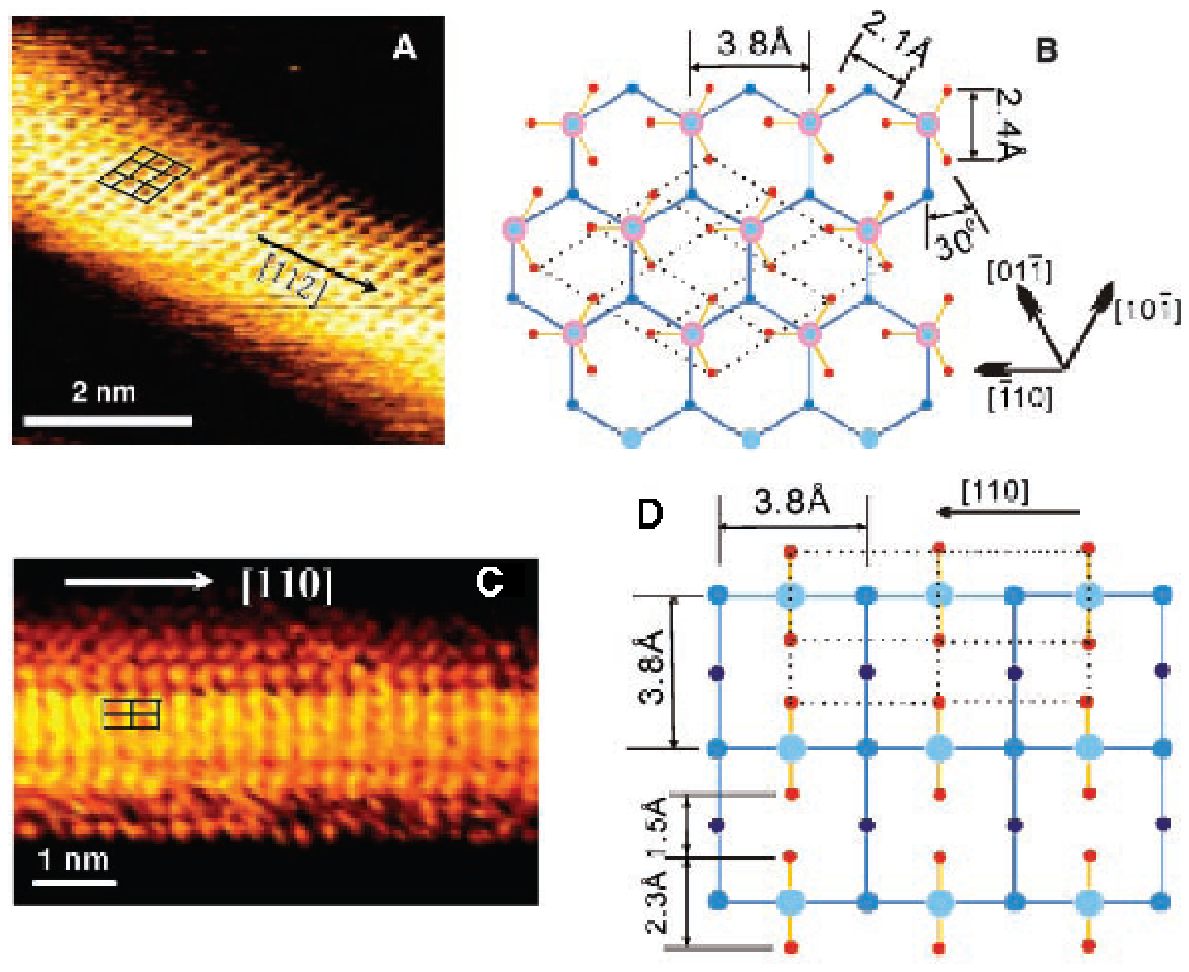, width=\linewidth, clip=, angle=0}
\vskip 0cm
\caption{(Color online) (a)~Constant-current scanning
         tunneling microscope (STM) image
         of the \{111\} facet of a SiNW grown along
         the \iiz\ direction; (b)~Schematic view of
         SiH$_3$ on Si(111) viewed along the \iii\
         direction; (c)~Constant-current STM image of
         the \{100\} facet of a SiNW grown along the
         \iio\ direction; (d)~Schematic view of the
         dihydride phase on Si (001). Red and large
         blue circles represent the H atoms and Si atoms,
         respectively. Small blue circles correspond to
         Si atoms on the underlying layers. The \iiz\ wire
         in panel (a), with a diameter of 1.3~nm, is the
         thinnest SiNW reported to date.
         Adapted from \textcite{MaScience03}. Reprinted with
         permission from AAAS.}
\label{fig:MaScience03fig1-2}
\end{figure}

Surface passivation in SiNWs mainly originates from two causes:
(i)~the growth of a thin layer of SiO$_2$ by thermal 
oxidation of silicon; (ii)~presence of hydrogen in the 
growth environment during the synthesis or HF attack of 
the oxidized wires after growth, a process yielding 
removal of the SiO$_2$ layer and H passivation. Hydrogen 
passivation is rather simple to model. If a sufficient amount of 
hydrogen is supplied the H atoms readily terminate each 
Si DB by forming a stable Si-H system. Passivation by 
oxidation is more complex. Thermal oxide is amorphous 
and then difficult to model at the nanoscale, because of 
the large amount of atoms required to describe the 
disordered phase. In the study of SiNWs, for most 
practical effects, hydrogen termination is a reasonable 
approximation to oxide passivation and this is the 
strategy adopted in most of the theoretical studies 
reviewed here. This approach is also justified by 
the fact that it is easy to remove the oxide layer after 
the growth and to induce H passivation by simply etching 
it with HF. This procedure is often followed~\cite{MaScience03,
WuNL04,RossPRL05,GuichardNL06,HeNatNano06,WangMatSciEng08} 
in order to work with {\em cleaner} structures where the 
passivation rely on an individual termination of the DBs,
rather than a less controllable and more defective oxide 
coverage~\cite{BaumerAPL04}. Furthermore, it has also 
proven to leave the morphology of the nanowire essentially 
intact, except for the removal of the oxide layer~\cite{ZhangPRB00}, 
allowing inspection of the underlying atomic scale structure 
(see Fig.~\ref{fig:MaScience03fig1-2}). Yet, more attention 
is likely to be devoted in the near future to the specific 
nature of SiO$_2$ passivation, beyond the simple models 
considered so far~\cite{AvramovPRB07}.

The passivation has a crucial effect on the electronic 
structure of the wires and it is essential to provide 
the wires with predictable band gap widths and an invariably 
semiconducting character. We will discuss these topics in 
detail in Section~\ref{sec:elec}.

Surface passivation has also an important effect
on the structural arrangements of SiNWs.
Besides preventing complex reconstructions, it
also influences the structure of the sub-surface
and innermost part of the wires. H-passivated
SiNWs grown along different orientations have
been found to maintain remarkably the bulk
symmetry (see Fig.~\ref{fig:SinghNL06fig1mod}), 
with negligible deviations of the Si-Si bond lengths; 
the deviation increases close to the surface, depending 
on the level of surface rearrangement~\cite{VoPRB06}. 
The limit case in this sense are pristine wires, where 
the absence of passivation results in major surface 
rearrangements and large deviations of the Si-Si bond 
length also in the wire core~\cite{KagimuraPRL05}.

An interesting path to the determination of the structure
of H-passivated \iio\ SiNWs has been proposed by 
\textcite{ChanNL06}. Their optimization procedure is based 
on a genetic algorithm. With this method, in principle 
suitable for any other growth orientation, they identified 
a pool of {\em magic} structures~\footnote{{\em Magic} 
structures is used in this context to refer to distinct 
types of wire configurations with low formation energies 
that emerge as the number of atoms per length is 
increased~\cite{ChanNL08}.} for \iio\ wires. Although 
some of them have not been observed experimentally, 
their hexagonal structure provided a good agreement with 
the STM image of the wire facet of \textcite{MaScience03}.

A more systematic approach was followed by 
\textcite{ZhangJCP05} in a study analogous to 
the one performed by \textcite{JustoPRB07} for
pristine nanowires.
They carried out a comprehensive study of
the possible low-index facets in H-passivated SiNWs grown 
along the \ioo, \iio, \iii, and \iiz\ axis. While many
choices are possible for \ioo, \iio, and \iii\ wires,
they showed that there is only one low index configuration 
--with two \{111\} and two \{110\} facets-- for \iiz\ 
wires. It is suggested that this would ease the controlled 
growth with a predetermined cross-section and could have 
important consequences on the engineering of devices 
based on SiNWs. In Section~\ref{sub:pass_el}, however, 
we will see that it has been recently suggested that the 
exact cross-section shape is less important than other 
parameters --such as the effective diameter and the 
surface-to-volume ratio-- when it comes to determining 
the electronic properties of SiNWs. 

Another important aspect to consider in H-passivated
SiNWs is the surface structure of the hydrogenated 
facets. This issue has been tackled by \textcite{VoPRB06},
where a systematic study of the effects of varying the
diameter and the growth direction has on the structure
of the hydrogenated surfaces of \ioo\, \iio\ and \iii\ SiNWs.
In particular, they studied the relative stability of
symmetric SiH$_2$ dihydrides, canted SiH$_2$ dihydrides
and a (2$\times$1) surface reconstruction (where first
reconstruction is allowed and then passivation occur),
see Fig.~\ref{fig:VoPRB06}. They
deliberately chose simple, round cross-sections, as
their scope was focusing on the atomic scale structure
of the facet. Their wires were constructed selecting all
the atoms falling inside a virtual cylinder placed in
bulk silicon, in such a way that the facets approximated
a circular cross-section. This procedure agrees with the
smoothness prescription described above, which --more
importantly-- seems also to be confirmed by the
experiments~\cite{MaScience03}.
They found that, in agreement with bulk Si(100) 
surfaces~\cite{NorthrupPRB91}, the canted dihydride 
surface is more stable than the symmetric dihydride 
structure, because canting allows a larger H-H separation. 
Additionally, faceting confers an increased stability to the 
canted dihydride surface, because at the facets' edges 
the SiH$_2$ groups are free to rotate. Relief of the
surface strain through bending as an additional
mechanism has been explored by \textcite{ZdetsisAPL07}.

A perhaps more flagrant effect of
the surface induced strain is the fact that the axial 
lattice parameter of thin SiNWs is in general different
from bulk Si. \textcite{NgPRB07} reported contraction
along the wire axis for \ioo\, \iii\ and \iiz\ SiNWs,
and elongation for \iio\ growth orientation. 


Other types of surface passivation --including 
OH~\cite{NolanNL07, NgPRB07,AradiPRB07}, 
NH$_2$~\cite{NolanNL07}, F~\cite{NgPRB07}, 
or Br, Cl, and I~\cite{LeuPRB06}-- have been 
considered. While changing the passivation has 
a limited effect the structural properties of the 
nanowire, it can affect in a more significant way 
the electronic band structure. We will come back on 
this topic in Section~\ref{sub:surf_struct}.

As a conclusive remark one should notice that, despite 
the intensive research carried out to find the equilibrium 
shapes for the different growth orientations --proposing 
structures that range from fullerene-like~\cite{MarsenPRB99}
 to star-shaped~\cite{SorokinPRB08}-- in most cases the 
experimentally observed cross-sections of passivated 
SiNWs are deceptively simple (see Fig.~\ref{fig:MaScience03fig1-2} 
and \ref{fig:WuNL04fig4}), whereas unpassivated SiNWs have 
never been reported.
Furthermore, as we shall see in Section~\ref{sec:elec},
although the cross-section shape has captured great 
attention and has been the object of many studies, 
in realistic, passivated wires the growth orientation 
and the average diameter turned out to have a more significant 
impact on the electronic properties of SiNWs.

\subsection{Mechanical properties of nanowires}
\label{sub:mech}

If one {\em carves out} of bulk Si a rod-shaped system 
like a nanowire, there is no apparent reason to expect
an enhanced stiffness, while the larger surface-to-volume 
ratio is rather suspected to be detrimental. A simple 
way of understanding these effects is that there is a 
layer of material at the surface and edges whose mechanical 
properties differ from those of the bulk including different 
elastic moduli and eigenstrains. 

\begin{figure}
\centering
\epsfig{file=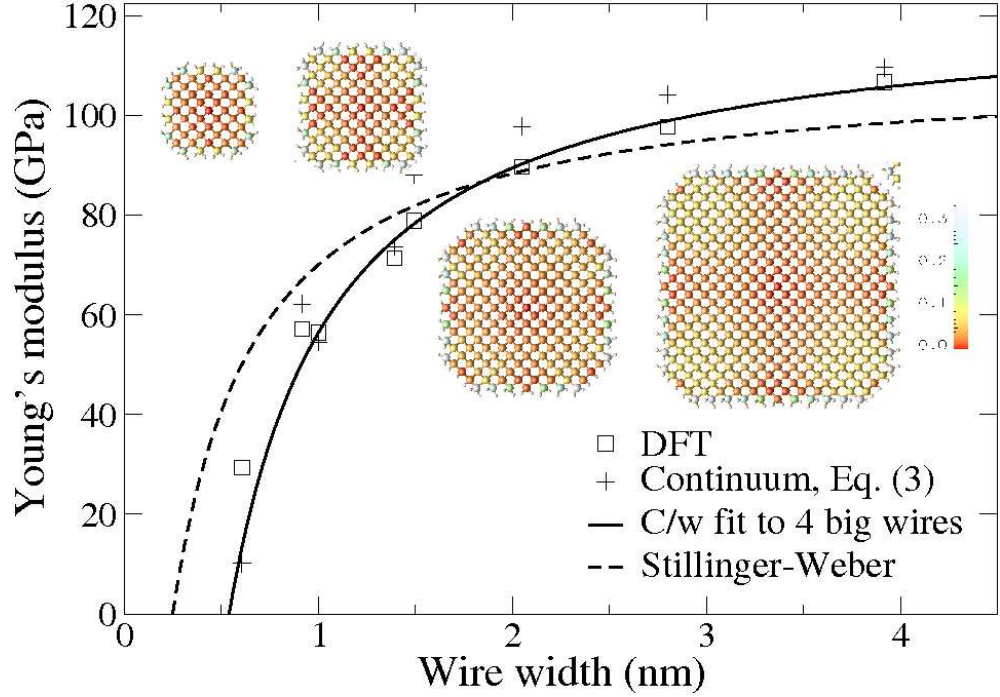, width=0.80\linewidth, clip=, angle=0}
\vskip 0cm
\caption{(Color online) Young's modulus calculated within
         DFT as a function of wire size. For comparison
         values of continuum formula are also plotted.
         The solid
         curve $E=E_{bulk}^{DFT} - C/w$, where $w$ is
         the width of the wire and  $C=66.11$~GPa/nm,
         is the best fit to a pure surface area to volume
         size dependence. (Insets) Cross sections of
         some of the SiNWs studied, where each Si atom
         is colored corresponding to its transverse
         relaxation in \AA. The widths of wires are
         (a)~1.49, (b)~2.05, (c)~2.80, and (d)~3.92 nm.
         Adapted from \textcite{LeePRB07}. \vskip 50pt}
\label{fig:LeePRB07fig1-3}
\end{figure}

These intuitive ideas have been rigorously tested by 
\textcite{LeePRB07}, by means of an exhaustive study of 
\ioo\ SiNWs with increasing diameters. 
They
calculated the Young's modulus, finding that it softens 
from the bulk value as the surface-to-volume ratio
increases, going through a steep decrease around 
2-2.5~nm diameter (see Fig.~\ref{fig:LeePRB07fig1-3}). 
They showed that the origin of this behavior 
is the compressive surface stress. To get a better insight 
into these atomic scale mechanisms 
the Young's modulus can be decomposed into a core 
(Si core atoms) and a surface contribution (Si surface 
atoms, H-H and Si-H systems). This decomposition allows 
highlighting the insensitivity to the facet ratio, as the 
contributions to the Young's modulus that are strongly 
facet dependent are very small. These first-principles 
results are in good agreement with empirical atomistic 
potentials and continuum techniques~\cite{LeePRB07b}, 
unless for the smallest wires where these simplified 
approaches fail (see Fig.~\ref{fig:LeePRB07fig1-3}).

The Young's modulus, as many other properties
of ultra-thin SiNWs reviewed in this
paper, is strongly anisotropic.
\textcite{MaCPL08} extended the study of the
stiffness vs. diameter to wires grown
along the \iio, \iii, and \iiz\ orientations.
While their results are in good agreement with
those of \textcite{LeePRB07} for \ioo\ wires,
they showed that wires of similar diameter,
but with different orientations, differ 
considerably. In particular they found the 
highest values for \iio\ SiNWs, while \ioo\ SiNWs 
give the lowest values. These results are in
good agreement with the work of \textcite{LeuPRB08}
where the Poisson ratio is also considered.

We note that to calculate the Young's modulus a
definition of the cross-sectional area must be
assumed and it is not univocal. We will run
into a similar problem concerning the definition
of the wire diameter in Section~\ref{sub:quantum}
when dealing with quantum confinement.
\textcite{LeePRB07} took the area bounded by the
outermost atoms, i.e. the passivating H atoms, 
\textcite{MaCPL08} made a similar choice, but
excluding the H atoms. Aware of this 
degree of arbitrariness in the possible choices,
\textcite{LeuPRB08} studied the variation of the
calculated mechanical properties as a function
of the uncertainty $\delta r_0$ in the estimation 
of the radius $r_0$. The error in the Young's modulus 
is $2\delta r_0/r_0$ and goes to zero in the limit
of large wires ($r_0 \to \infty$). The Poisson
ratio, on the other hand, is much more sensitive, 
because the error is $-\delta r_0/r_0+(\delta r-
\delta r_0)/(r-r_0)$, $r$ being the radius at a strain 
$\epsilon$; while the first term vanishes for large 
radius, the other is always present and can be 
significant as $(r-r_0)$ is typically small.

Bending has received comparatively less attention,
although a few experimental measurements have been 
reported~\cite{Tabib-AzarAPL05,HoffmannNL06,HsinAdMat08,ZhengNL09}. 
This is probably due to the difficulty of studying a bended nanowire 
within atomistic simulations that normally relies on periodic 
boundary conditions. The fabrication of ingenious mechanical 
structures with enhanced elastic properties suggests that 
this could be a promising research direction~\cite{SanPauloNL07}.

Beyond elastic deformation, materials undergo 
non reversible, plastic deformation which 
directly precede fracture. This regime has been 
studied for \ioo\ pristine SiNWs~\cite{JustoPRB07} 
and for \iii\ and cage-like SiNWs~\cite{MenonPRB04}, 
using two different interatomic potentials~\cite{JustoPRB98,
StillingerPRB85}.
At small strains $\epsilon$ the stress increases linearly,
as expected in the elastic regime, while at larger 
deformation the plastic behavior appears until the 
fracture occurs at $\epsilon \sim$~0.10, with a good 
agreement between the two different models.
Experimentally, however, the fracture is much delayed
with respect to theoretical predictions and the wire
breaks at $\epsilon \sim$~0.25~\cite{KizukaPRB05}.
It should be mentioned that both these theoretical 
studies considered pristine nanowires, while the wires 
in the experiments are coated with a thin layer of oxide,
thus it is difficult to make a rigorous comparison. At
the same time it is not surprising that a different
surface treatment can produce a noticeable difference in
the mechanic response, because it is just at the surface 
that the nanoscale signature emerges.

Correlating structural deformations and changes 
in the electronic properties is an active field of 
research~\cite{RuraliPRB08} and the use of strain 
to enhance carrier mobility has been investigated~\cite{LeuPRB08,
HongNL08,HuangJPCC08}. 
Furthermore, a giant piezoresistance effect --the 
application of a strain to a crystal that results in 
a change in the electrical resistance-- has been 
reported recently~\cite{HeNatNano06}. The underlying 
atomic scale mechanism is still poorly understood, 
however, and the attempts made so far proved to be 
somehow elusive~\cite{CaoPRB07,RoweNatNano08}.

\section{Electronic properties}
\label{sec:elec}

\subsection{Pristine Nanowires}
\label{sub:pure}

The reason for the attention devoted to geometrical
features such as the growth orientation, the faceting
arrangement, and the surface structure, is that they 
are crucial when it comes to the electronic properties of 
the nanowire. Clearly, the thinner is the nanowire,
the more it is sensitive to the structure details,
as in the limit of very large diameter --no matter 
which is its crystal axis or cross-section shape-- 
its properties converge to those of bulk Si. 

As we discussed in the previous section, pristine
nanowires turned out to have a limited relevance,
at least to date, because experimentally grown 
SiNWs are always passivated. However, the study
of bare, unpassivated wires is still interesting
for two reasons: (a)~it leads to the important 
conclusion that passivation is essential to obtain 
nanowires with predictable and easy to control 
electrical properties; (b)~it sheds a light on some
atomic scale mechanisms of high fundamental interest.


An interesting example in this sense is the electronic
structure of \ioo\ SiNWs with \{100\} facets. While other 
facets, like the \{111\} facets, have an electronic 
structure similar to the corresponding infinite 
surface~\cite{PandeyPRL81,RuraliPRB06},
\{100\} facets can be very 
different. In the Si(100) surface each surface atom has 
two DBs. The surface is known to reduce its energy 
by forming dimers, thus halving the number of 
DBs~\cite{ChadiPRL79}. The reconstruction of \{100\} 
facets follows the same pattern, but wires dominated by 
such facets have been reported to be metallic.
\textcite{RuraliPRL05} showed that thin \ioo\ SiNWs
sustain two different reconstructions of the \{100\}
facet that turn the wire metallic or semimetallic, 
in agreement with what previously suggested by 
\textcite{Ismail-BeigiPRB98}. The metallic behavior 
can be ascribed to a modified coordination 
of the \{100\} facet atoms, leading to a distortion 
of the surface dimers, with respect to the Si(100) 
surface~\cite{RuraliPRB06}. The metallicity rapidly 
vanishes as the diameter is increased and the facets 
recover the coordination and the semiconducting electronic 
structure of the Si(100) surface. 

The \ioo\ wires with 
sharp corners studied by \textcite{CaoPRL06} can be 
metallic too. Interestingly, the edge metallic states decay 
slower with the diameter compared to the facet metallic 
states. Consequently, wires thicker than those considered
by \textcite{RuraliPRL05}, where edges were absent, can 
still be metallic.
In both cases the metallic states are related directly or indirectly
with the edges --in one case purely edge states, in the other dimer
rows with an altered coordination near the edges; as the
wire size increases the relative number of atoms at the edges
decreases rapidly and the facet recovers the semiconducting character
of the infinite surface.

Besides the fundamental interest of these findings
--in one case a metallicity driven by the finite size
of the facet, in the other sustained by the edges--
it is clear that such wires are not desirable for
electronics application. On the one hand one
wants to work with semiconducting systems; on the
other hand, although some application can be envisaged 
for metallic SiNWs, e.g. interconnects, the
metallicity should be much more robust, so that
it is not destroyed by small variations of the diameter 
and does not depend critically on the atomic scale 
structure of the wire.


A comprehensive study of the surface reconstruction
and electronic structure of pristine \iio\ wires has 
been carried out by \textcite{SinghNL05}. The cross-section
chosen for these wires is such that they have \{100\} and 
\{110\} facets, at variance with the SiNWs of 
\textcite{RuraliPRB06} which have \{111\} and \{110\}. 
This variation results in significant structural differences,
because of the comparatively larger readjustment of \{100\}
facets, which involve the formation of surface dimers~\cite{ChadiPRL79}
and therefore a noticeable reconstruction; in \{110\} and \{111\} facets,
on the other hand, no new bonds are formed and the overall
reorganization of surface atoms is moderate. These \iio\
SiNWs turned out to be indirect band gap semiconductors,
with the states of the top of the valence band and the 
states of the bottom of the conduction band originating at 
different facets. Yet, it should be noted that a metallic 
reconstruction for \iio\ SiNWs have been reported by 
\textcite{Fernandez-SerraPRL06}. Once again, small 
variations of the atomic scale structure or of the 
cross-section can result in major changes in the electronic 
structure.

Contrarily to what we shall discuss in 
Section~\ref{sub:quantum} concerning quantum confinement,
here the thinner is the wire, the smaller is the
effective band gap. This looks like a general feature of
pristine, unpassivated nanowires, where band gaps
are smaller than in bulk~\cite{RuraliPRB05,RuraliPRB06}.
In thin wires the surface-to-volume ratio is larger and 
surface states, which often lie in the gap, dominate
the electronic structure and result in an effective 
narrowing of the energy gap.

\subsection{Passivated nanowires}
\label{sub:pass_el}

\subsubsection{Band structure and band gap anisotropy}

\begin{figure}
\centering
\epsfig{file=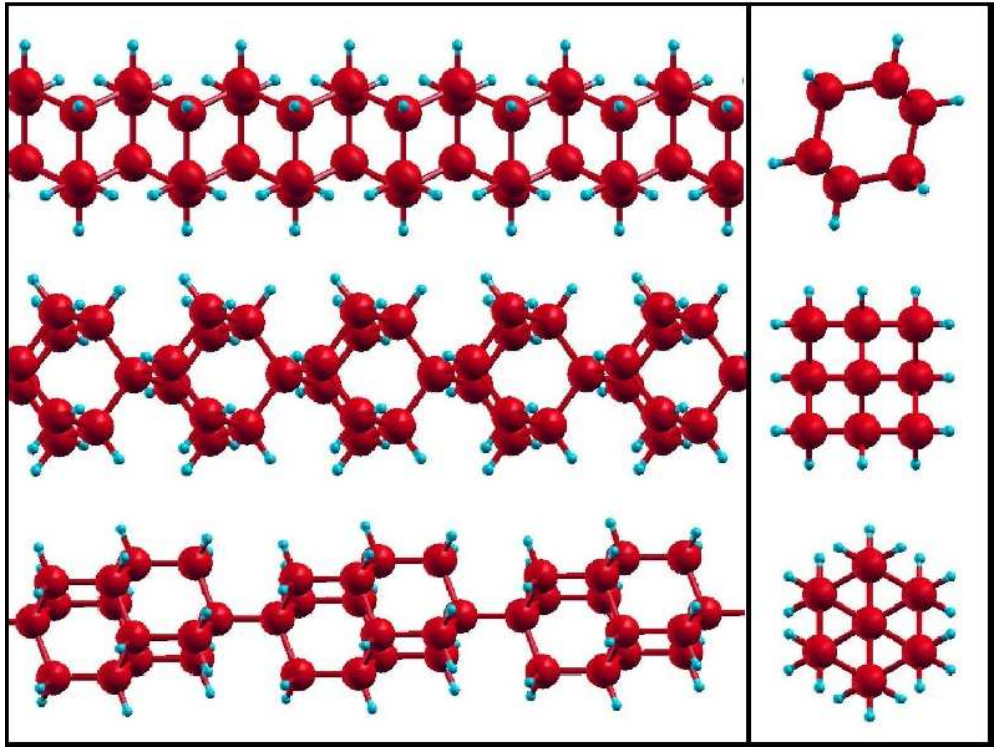, width=0.80\linewidth, clip=, angle=0}
\vskip 0cm
\caption{(Color online) Geometrical structures of 0.4~nm Ge 
         nanowires along the \iio\ (top), \iii\ (middle), and \ioo\ 
         (bottom) directions shown from the side (left) and from the 
         top (right). Large spheres represent Ge atoms; small spheres 
         are hydrogen atoms used to saturate the dangling bonds.
         Adapted from \textcite{BrunoPRB05}.}
\label{fig:BrunoPRB05}
\end{figure}

We have already mentioned a few times that surface
passivation is required to obtain ultra-thin nanowires
that are semiconducting and have a predictable and 
controllable band gap. Notwithstanding, the electronic 
structure of the nanowires still depends on the growth 
orientation, on the cross-section shape and on the diameter. 
The band gap is strongly anisotropic~\cite{ZhaoPRL04,
LeuPRB06,SinghNL06,NiquetPRB06,NgPRB07,VoPRB06,YanPRB07,
RuraliPRB07,LeuPRB08} and, for wires of comparable 
diameters, it follows the ordering 
\begin{equation}
E_g^{\langle 100 \rangle} > E_g^{\langle 111 \rangle} \sim 
E_g^{\langle 112 \rangle} > E_g^{\langle 110 \rangle},
\end{equation}
with the orientation effect still sizeable up to 3~nm 
diameter~\cite{NgPRB07}. The band gap of \iiz\ wires
is of the same order of \iii\ wires, though it has been 
reported to be slightly larger~\cite{LeuPRB06,RuraliPRB07} 
or slightly smaller~\cite{NgPRB07}.
This anisotropy has been qualitatively tracked back to the
different geometrical structure of the wires in the
\ioo, \iii, and \iio\ directions~\cite{BrunoPRB05,BrunoSurfSci07}.
While the \ioo\ and \iii\ wires appear as a collection 
of small clusters connected along the axis, the
\iio\ wires resemble a linear chain (see Fig.~\ref{fig:BrunoPRB05}
where the case of Ge nanowires is shown). Therefore, one 
expects that quantum confinement effects are larger in 
the \ioo\ and \iii\ wires, because of their quasi zero-dimensionality,
with respect to the \iio\ wires.

Bulk Si has an indirect band gap, with the valence band
maximum at the $\Gamma$ point and the conduction minimum 
at about 85\% along the $\Gamma$ to X direction,
and a phonon is required to conserve the momentum in any
electronic transition.
Remarkably, however, SiNWs grown along most of the 
crystallographic orientations have a direct band gap, 
meaning that the maximum of the valence band and
the minimum of the conduction band occur at the same
point in {\em k}-space. This property has allowed to 
envisage the use of SiNWs as optically active materials 
for photonics applications~\cite{CanhamAPL90,GuichardNL06}.

In \ioo\ SiNWs, the confinement plane contain four 
of the six equivalent conduction band valleys.
These minima at $\pm y$ and $\pm z$ are then
projected onto $\Gamma$ due to band folding, thus 
resulting in a direct band gap.
When the axis is along a lower symmetry direction
the confinement plane cannot contain four
conduction band valleys and it will contain
at most two.
This is the case of \iio\ SiNWs.
The minima at 
$\pm z$ are projected onto $\Gamma$ again. Now,
both the large and the small
masses appear in the confinement plane, with the larger longitudinal
mass being the relevant effective mass for describing
the confinement effect in the cross-section plane. On the
other hand, the four remaining minima will be projected to a
point between $\Gamma$ and the zone boundary Z, 
with the effective
mass on the confinement plane being a value between
the longitudinal and transverse masses. Therefore, the
conduction-band edge at $\Gamma$ is expected to have a smaller
upward shift induced by confinement and the band gap becomes
direct~\cite{YanPRB07}.

Although the projection along the \iii\ axis
would lead to an indirect band gap, the thinnest 
\iii\ SiNWs have a direct band gap~\cite{ZhaoPRL04,
VoPRB06,RuraliPRB07}.
One should bear in mind that, besides the
band folding arguments given above, the effective 
masses play an important role. In the quantum
confinement regime (see below, Section~\ref{sub:quantum})
the conduction band states are shifted upward,
the smaller the diameter, the larger the shift.
Nonetheless, the magnitude of this energy shift 
is different for each {\em k}-point of the band structure
and depends on the effective mass. In bulk Si the 
effective mass at $\Gamma$ is heavier than at $X$ 
or $L$. Hence, upon confinement, one expects the 
conduction band energy at $X$ and $L$ to increase 
more than at $\Gamma$. This simple considerations
based on effective mass theory (EMT) describe well
the transition from direct to indirect band gap
experienced by \iii\ SiNWs that occur around 2~nm:
as the diameter increases the quantum confinement
effect vanishes (see Section~\ref{sub:quantum}),
$\Gamma$ and $X$ and $L$ points are not shifted
and the gap remains indirect. Among the studied orientations 
\iiz\ SiNWs are the only ones that have an indirect band gap 
also for the thinnest diameters~\cite{ScheelPSSb05,
NgPRB07,RuraliPRB07,AradiPRB07,LuAPL08,HuangJPCC08}.

Although the band gap is highly anisotropic and,
as we shall see in the next section, strongly dependent
on the wire diameter, it is very interesting to observe
that it is rather insensitive to the shape of the
cross-section. \textcite{NgPRB07} have studied the effect 
of the variation of the cross-section in thin \iio\ SiNW,
generating 13 different cross-sections obtained by
modifications of a reference 1~nm wire. They found
that the band gap is practically constant and changes
are within 0.09~eV. 
Later, it was demonstrated that wires of even utterly
different cross-sections can have the same band gap,
provided that their surface-to-volume ratio is the 
same~\cite{YaoNL08}. The effect of the surface-to-volume
ratio on the band gap can be described by the universal 
expression
\begin{equation}
E_{gap} = E^{bulk}_{gap} + aS
\end{equation} 
where $E^{bulk}_{gap}$ is the gap of bulk Si, $a$ is an 
adjustable parameter and $S$ is the surface-to-volume (in 
nm$^{-1}$). 

\subsubsection{Quantum confinement}
\label{sub:quantum}

One of the most intriguing physical effect that arise 
in confined systems like SiNWs is the so-called
{\em quantum confinement}. Such a regime is conveniently
described through the particle-in-a-box model system
in most quantum mechanics text books~\cite{BransdenJoachain}.
The simplified situation considered is an infinite 
potential well where the motion of the particles is
restricted to be in the direction of the confinement.
As the motion of the particles is restricted, their 
kinetic energy increases and it is readily shown
that the eigenstate energies are given by the following
relation:
\begin{equation}
E_n = \frac{\hbar^2 n^2 \pi^2}{2m^* d^2}
\label{eq:qc}
\end{equation}
where $m^*$ is the effective mass and $d$ the width of 
the potential well.
According to Eq.~\ref{eq:qc}, not only the energy levels, 
but also the spacing between them increases as the 
confinement becomes more pronounced, i.e. the smaller 
is $d$. 
Quantum confinement has a 
critical impact on semiconductors because it affects 
directly their most important electronic property: the 
energy band gap.

Semiconducting nanowires provide a very good approximation
of the model situation described above. Clearly, the potential
well is not infinitely deep and realistic wire cross-sections
like those described in Sections~\ref{sub:pristine} and 
\ref{sub:pass} are difficult
to describe analytically, thus there is a need for a detailed 
electronic structure modeling. 

The first experimental proofs of quantum confinement in 
nanostructured Si were reported in the pioneering works
of \textcite{CanhamAPL90} and \textcite{LehmannAPL91},
where a simple electrochemical etching process was used 
to create crystalline Si nanostructures with visible 
luminescence at room temperature. As TEM images revealed
later~\cite{CullisNature91}, the etched structures 
consisted of rather disordered 
bundles of nanowires, though it is interesting to note 
that ordered structures like those speculated in the 
first place~\cite{CanhamAPL90} have been recently 
proposed for the fabrication of ordered arrays of 
quantum wires~\cite{RuraliAPL07} and to achieve enhanced 
thermoelectric effect~\cite{LeeNL08}. 

\textcite{ReadPRL92} and \textcite{BudaPRL92} performed 
DFT calculations of the band gap upshifts in perfect 
H-terminated SiNWs as a function of wire thickness, 
modeling porous Si~\cite{CanhamAPL90} with 
rectangular columns oriented along the \ioo\ axis. 
Both these works showed that the fundamental gap
is direct at the $\Gamma$ point. This makes by itself
the probability of radiative recombination higher
than in bulk Si, since no phonon is required
in the electron-hole recombination process.
Unfortunately, as it is well-known, standard local 
and semi-local implementations of DFT fail to 
account quantitatively for the band gap of semiconductors 
and one must resort to self-energy corrections to the 
Kohn-Sham gap to obtain a good agreement with the 
experimental values. Yet, the trends are expected to be
qualitatively correct~\cite{WilliamsonPRL02} and 
\textcite{ReadPRL92} reported band gap upshift of up 
to 2~eV for wires of $\sim$~12~\AA\ diameter. They also 
shown that a 
generalization of Eq.~\ref{eq:qc} gives a good description
of the quantum confinement for wires wider than 23~\AA,
whereas thinner wires show significant deviations from this 
idealized EMT picture. In such a range \textcite{BudaPRL92}
showed that with the more realistic DFT potential the band 
gap scales as the inverse of the diameter $d$, rather than 
$1/d^2$ as predicted by particle-in-a-box arguments
where infinitely hard walls are assumed.
Subsequent studies that the interested reader could look 
at include the works by \textcite{OhnoPRL92}, 
\textcite{SandersPRB92}, \textcite{HybertsenPRB93}, 
\textcite{XiaPRB93}, \textcite{YehPRB94}, \textcite{SaittaPRB96}, 
\textcite{XiaPRB97}, and \textcite{OssiciniTSF97}.
 
A first step toward a quantitative evaluation of SiNW
band gaps in the quantum confinement regime was given
by \textcite{DelleyAPL95}, including a constant self-energy 
correction independent on the size. Namely, they increased 
all their calculated band gaps by 0.6~eV, the self-energy 
correction for bulk Si. They also shown that EMT can
predict with great accuracy the band gap of relatively
thin nanowires, provided that the potential well is not
assumed to be infinite. 

\begin{figure}
\centering
\epsfig{file=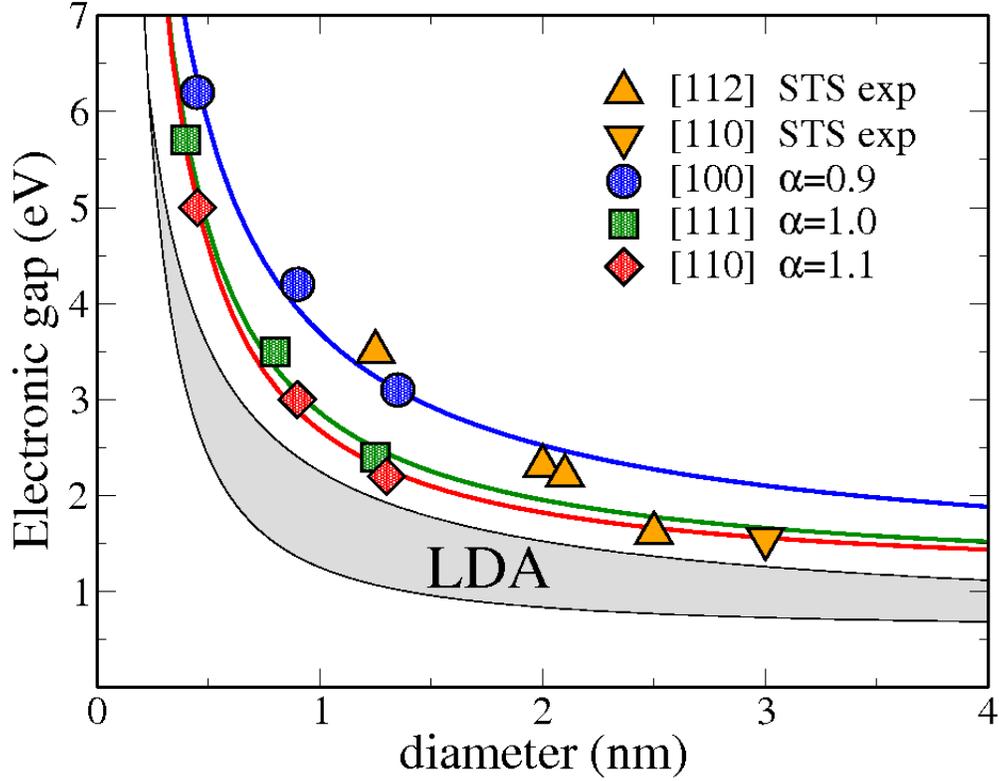, width=0.80\linewidth, clip=, angle=0}
\vskip 0cm
\caption{(Color online) Quasi-particle GW gaps for \ioo\
         (circles), \iii\ (squares), and \iio\ (diamonds)
         SiNWs as a function of wire size compared with
         experimental results (triangles) from scanning
         tunneling spectroscopy~\cite{MaScience03}.
         The gray region represents the LDA electronic gaps
         from \iio\ (bottom) to \ioo\ (top) wires. From
         \textcite{BrunoPRL07}.}
\label{fig:BrunoPRL07}
\end{figure}

The self-energy correction to the Local Density Approximation 
(LDA) or the Generalized Gradient Approximation (GGA) band-gap,
however, is expected to depend on the wire diameter
and on the growth orientation. \textcite{ZhaoPRL04}, 
\textcite{BrunoPRL07}, and \textcite{YanPRB07}
carried out calculations within the many-body 
perturbation method based on the GW 
approximation~\cite{AryasetiawanRPP98} for \ioo, \iio, 
and \iii\ SiNWs. They showed that the self-energy is 
indeed anisotropic and is larger for thinner wires. 
The dependence of the band gap on the wire diameter 
can be described as 
\begin{equation}
E_{gap} = E^{bulk}_{gap} + C \times (1/d)^\alpha,
\label{eq:gap}
\end{equation}
where $E^{bulk}_{gap}$ is the calculated band gap 
of bulk silicon, $d$ is the effective diameter of the 
wires, while $C$ and $\alpha$ are fitting parameters. 
This formula is derived within a simple particle-in-a-box 
effective mass approximation, where $\alpha = 2$ when
barrier height is infinite. The GW results can be fitted 
to this formula, yielding values of $\alpha$ ranging 
from 0.9 to 1.1, much lower than those expected within 
EMT and depending on the growth orientation (see 
Fig.~\ref{fig:BrunoPRL07}), so that the band gap and
the dielectric response are anisotropic~\cite{ZhaoPRL04,
BrunevalPRL05,BrunoPRL07}.

Although GW is in principle the best suited methodology to calculate 
the band gap in semiconductor systems, it suffers from the
serious inconvenience of a considerable computational load.
In the works of \textcite{ZhaoPRL04}, \textcite{BrunoPRL07} 
and \textcite{YanPRB07}, for instance, only relatively small 
SiNWs can be calculated directly and the band gaps of larger, 
more realistic wires are obtained by numerically fitting the 
available data to Eq.~\ref{eq:gap}. Furthermore, an 
alternative to many-body GW calculations is mandatory when
it comes to calculate doping levels, a task that requires large
computational cells. In the remainder of this section we 
discuss two possible approaches. 

A successful way to improve the DFT band gaps consists in 
using hybrid-functionals for the exchange-correlation
energy, where a certain amount of exact Hartree-Fock
exchange is mixed to conventional LDA/GGA functionals. 
The amount of Hartree-Fock exchange (typically 12-15\%)
is chosen to reproduce some parameters of the bulk
system (the band gap, among them), 
rather than being based on solid theoretical grounds.
Hence, strictly speaking, one cannot claim to solve 
the electronic structure {\em from first-principles}.
In such a theoretical framework the band gap of SiNWs
with diameters up to 3~nm can be calculated 
directly~\cite{NgPRB07,RuraliPRB07,AradiPRB07}. These 
results are important because they allow {\em direct} 
comparison
with the only experimental measurements available to
date~\cite{MaScience03}. A direct comparison of the
experimental data with GW calculations is not possible 
for two reasons: the diameters of the wires grown 
experimentally are larger than those that could be 
simulated and most of the available measurements are 
for for \iiz\ SiNWs, whose larger primitive cell 
precludes GW calculations even for the thinner wires. 

Alternatively, the band structure of nanowires can 
be calculated with a semiempirical tight-binding 
method, where the self-energy is obtained within a 
simpler semiclassical treatment of the image charge 
effects~\cite{NiquetPRB06}. This is a very powerful
method because, due its reduced computational load,
it allows calculating SiNWs with diameters up to
10~nm with good accuracy. As we will see in some 
more detail below (Section~\ref{sub:dielectric_doping}), 
a great advantage of this method is that it 
allows dealing with different dielectric surroundings, 
which is very important in systems with abrupt dielectric 
interfaces like nanowires~\cite{LiPRB08}. 

So, how should the band gap of SiNWs be calculated?
The accurate calculation of band gap is one of the most
challenging problems in semiconductor theoretical physics,
so it is not surprising that it is not easy to answer 
this question. GW calculations provide in principle 
the most accurate estimations. However, they are 
restricted to very thin SiNWs.
Semiempirical tight-binding, on the other hand, is a very
attractive choice for larger wires, especially
for those diameters where quantum confinement become 
small and gap broadening is dominated by dielectric 
mismatch effects~\cite{PereiraPRB09}. Hybrid-functional DFT calculations
are an interesting compromise for those wires that
are too large for GW (too many atoms in the primitive
cell) and too small for tight-binding (relaxation effects cannot be 
neglected and the use of a parametrization obtained for
bulk Si could be questioned). 
It is difficult to asses on the accuracy of each of
these methods, since the experimental measurement
of the energy gap of SiNWs is extremely challenging,
and only the data of \textcite{MaScience03} are
available to date. More experimental results are needed
to clarify this important point.

\subsubsection{Surface chemistry}
\label{sub:surf_struct}

We have seen above --and will see again below-- that 
many properties of SiNWs are determined by their large 
surface-to-volume ratio. Hence, it is natural that most 
of the exciting physics takes place at the wires' 
surface~\cite{KobayashiPRB04,ZhongNL06}.
In Section~\ref{sub:pure} we have seen, for instance, 
that wires bounded by facets derived by semiconducting 
surfaces can exhibit surface metallicity.
Passivated nanowires are more predictable, in this sense,
and it is just because they are always
semiconducting that they are expected to play an
important role in the next generation of electronic
devices. Yet, the surface has a relevant role
that merits some considerations.

An important case is that of chemical sensors where
the adsorption of a molecule yields measurable variations
of the electrical conductance~\cite{CuiScience01b,
BlasePRL08}. Upon adsorption, the
molecular orbitals can hybridize with the wire states,
resulting in a sizeable modifications of its electronic
structure. Whether the effectiveness of this process 
depends on the facet where adsorption takes place 
is the topic addressed by \textcite{LeaoNL07}.
They studied the sensitivity of 
different facets of a \iio\ SiNW, showing the existence 
of a specific relation between the way surface atoms are 
bonded to core atoms and the relative contribution
of these surface atoms to band edge states. These 
observations are important concerning the optimal design 
of those chemical sensors where the adsorption of a 
molecule directly modify the wire transmission. A broader 
class of sensors, however, seems to work on simpler basis. 
The dipole induced by molecule adsorption can act as a 
gate voltage, opening or closing the conductive channel 
in a field-effect transistor set-up.

In the quantum confinement regime the band gap width
depend critically on the diameter. The possibility
of controlling the band gap width is tremendously
attractive for optoelectronics applications: not
only SiNWs can have a direct band gap, which {\em per
se} increases the optical efficiency, but its width can 
in principle be tuned. It is not difficult to imagine, 
however, that controlling the wire diameter with tolerances
within 1-3~nm is a more than challenging task.
A simpler route to band gap tuning is controlling 
the chemical composition and the coverage density
of the wire surface.
Halogens such as Cl, Br, and I can be used as surface
passivation agents instead of H and, while not altering 
the semiconducting character of the wires, they result 
in a significant shrinking of band gap~\cite{LeuPRB06}.
The strongest reduction of the band gap is provided by I,
followed by Br and Cl, in the opposite order of the
bonding strength of these species and SiNWs. 
Interestingly, the surface coverage is a further
degree of freedom and one can span all the band gap
values between a H- and halogen-passivated wire
by varying the H:halogen ratio. 
Also, increasing the halogen surface concentration 
the band edge states, concentrated in the wire core
in presence of H-passivation, progressively spread to the surface.

Analogous results have been reported for OH and 
NH$_2$~\cite{NolanNL07,AradiPRB07}. It should be noted
that the passivating species do not contribute significantly 
to the states close to the band edges, so that the
reduction of the gap is not caused by the introduction of 
additional bands. It rather comes from the hybridization
of the valence band states with the frontier orbitals of 
the different passivating functional groups that cause 
a significant band gap reduction relative to H-passivated 
wires. 

These results indicate that the band gap width in 
SiNWs can be tailored not only by controlling the wire 
diameter, but also by an appropriate choice of the surface 
termination.

\subsection{Doped and defective nanowires}

Semiconductors are privileged materials for electronics 
applications because their resistivities can be can be 
varied by design with great control~\footnote{Conventionally, 
materials with resistivities less than about 10$^{-2}\; \Omega$ 
cm are considered {\em conductors} and materials with 
resistivities greater than about 10$^5 \; \Omega$ cm are 
considered {\em insulators}~\cite{MullerKamins}.}.
Equally important, they can be designed to conduct one 
of two types of carriers: electrons and holes. These two 
features are the core of device design, which relies on 
the interaction of adjacent semiconductors with different 
densities of electrons and holes. The most efficient way 
to control the carrier density is {\em doping} the 
semiconductor, that is incorporating substitutional 
impurities ({\em dopants}) in the lattice~\cite{MullerKamins}.

In the simplest model, a substitutional defect with minor 
relaxation effects forms four bonds with the neighbor 
atoms in the Si crystal. For a group-V element, such as 
P or As, the fifth valence electron is not covalently 
bonded to near neighbors and it is only weakly bonded by 
the excess positive charge of the impurity nucleus. Hence, 
a small amount of energy is required to break this weak 
interaction and this electron is free to wander about 
the crystal and contribute to conduction. These impurities
are called {\em donors} because they {\em donate} an electron;
analogous arguments apply to group-III elements which are
{\em acceptors}.

There are at least two reasons that make the physics of 
impurities in nanowires different with respect to bulk 
systems: (i)~the lattice sites are no longer equivalent 
in the direction of confinement and (ii)~in the quantum 
confined regime all the eigenstates, including those 
associated to defects, are shifted in energy with
important consequences on the dopant activation.
Below, we discuss in some detail these and other topics
relevant to dopant efficiency. We conclude the section
with a generalization of the formation energy for
defects in nanowires.

\subsubsection{Surface segregation, surface traps and dopant aggregation}
\label{sub:surf_seg}

In bulk Si all the lattice sites are equivalent. In a 
nanowire this is true only for the axial direction, 
because the lateral confinement breaks the translational 
symmetry. In other words, given one arbitrary Si atom 
in the nanowire, as one moves along the wire axis he 
finds an infinite number of equivalent atoms, whereas 
as one moves along the radial direction he finds a finite 
number of nonequivalent atoms. Therefore, the 
substitutional defects at these non-equivalent sites
will have in general different formation energies and 
doping levels. 
 
This problem was first tackled by \textcite{Fernandez-SerraPRL06},
who studied B and P substitutional in \ioo\ and \iio\ 
SiNWs. They revealed a tendency to surface segregation 
of these impurities, which means that substitution close 
to the surface is energetically cheaper than
substitution in the innermost part of the wire. The 
effect is especially pronounced in presence of DB defects, 
so that the dopant impurities are 
effectively trapped by these surface defects. Most 
importantly, the dopant-DB complex is electrically 
inactive, reducing the carrier concentration at room 
temperature.

The segregation energy of P is significantly larger 
than that of B. This means that for the same impurity 
concentration, a much larger fraction of P atoms will 
be captured and neutralized by surface traps, resulting 
in a lower conductance. This is in agreement with
the experimental results~\cite{CuiJPCB00,YuJPCB00} 
where, for similar doping levels, B-doped SiNWs present 
a lower resistance than P-doped ones.
Similar studies have been carried out for \iio\ SiNWs 
of different diameters~\cite{PeelaersNL06,LeaoNL08}.
They showed that B and P
prefer to sit at edge or near edge sites (the most
external lattice sites with all-Si nearest
neighbors), depending on the surface facet and
on the atomic surface structure~\cite{LeaoNL08}.

In perfectly passivated wires the surface segregation 
originates from a simple {\em relaxation effect}. At these edge 
and near edge positions it is easier to release the strain 
induced by the substitution, whereas in the center of the 
wire the allowed relaxation is smaller due to the constraint 
of the surrounding Si lattice. In surface defective wires, 
on the other hand, the presence of DBs greatly enhances 
the tendency to surface segregation, with the impurity 
atoms moving at surface sites. Here the driving 
force to surface segregation, yielding a much more 
sizeable effect, is an {\em electronic effect}: the 
formation of a stable dopant-DB complex.

The surface trapping of dopants has a dramatic impact for 
two reasons: (a)~the impurities trapped at the surface 
are deep-state defects and are electrically inactive, 
thus not contributing to the carrier concentration at 
room temperature; (b)~due to their large surface-to-volume
ratio and considering a typical dopant concentration of 
5$\times 10^{18}$~cm$^{-3}$ and an estimate of of 
$10^{12}$~cm$^{-2}$ interface DB defects, one finds 
that for wires of less than 4~nm diameter there are 
always enough DBs to trap all the dopants~\cite{PeelaersNL06}.
Luckily, the difference in formation energies
among surface and core substitutional sites has
been shown to decrease rapidly as the diameters
grow larger~\cite{LeaoNL08}. Hence, there is a
twofold benefit in enlarging the wire diameter:
on the one hand, the surface-to-volume ratio
decreases, and so does the density of DBs
with respect to the dopants concentration; on the
other hand, the trapping efficiency of these
reduced density of surface defects is lower, 
as the formation energies of core substitutionals
and dopant-DB complexes become of the same order.
\textcite{LeaoNL08} estimate that the populations 
of core and surface dopants will be similar
for wires with diameters of 3~nm or more.

The importance of surface impurities has been further
highlighted by \textcite{DurgunPRB07}. They
considered various impurities such as Al, Ga, C,
Si, Ge, N, P, As, Te, and Pt, focusing on adatoms
configurations, rather than on substitutionals. 
They found that the energetically most favorable
adsorption site of the six considered depends on
the group of the Periodic Table that the impurity 
belongs to. All the configurations studied, however,
give rise to deep state in the gap and are not
viable choices as active dopants.

Another source of dopant deactivation is the formation
of electrically inactive dopant complexes. Two nearest 
neighbor dopants can form a bound state, so that the 
weakly bonded electrons that contributed to the conductance 
are now participating in the dopant-dopant bond. Yet, this 
energy gain has to compete with the energy cost that 
results from the strain accumulated around the dopant-pair 
defect. This strain is more easily released in nanowires 
than in bulk, due their large surface-to-volume ratio. 
\textcite{MoonNL08} reported a high stability of P pairs, 
which increases as the wire diameter is reduced. 
Interestingly, this is not the case of B. When two B 
atoms occupy nearest neighbor sites the lattice undergoes 
a significant relaxation, the B impurities move far apart 
and assume a planar threefold coordinated configuration. 
This is possible because, unlike P, B can present either 
sp$^2$ or sp$^3$ hybridization. Again, {\em p}-type 
doping seems to be more robust than {\em n}-type doping, 
at least as far as one considers B and P.
A similar mechanism 
leads to mutual passivation when both a B and a P impurity 
are present at the same time. Besides the obvious 
compensation of having an {\em n-} and a {\em p-}type 
dopant, \textcite{PeelaersNL06} showed that these 
impurities favor aggregation at the wire surface.
Also, the use of B/P co-doping has been proposed to reduce 
in a controllable way the band gap~\cite{IoriSuper08}.

\subsubsection{Quantum confinement}
\label{sub:quantum_doping}

In Section~\ref{sub:quantum} we have seen how one of the most 
important quantities of a semiconductor, its band gap,
depends critically on the wire diameter in the regime of
quantum confinement. This is a very important parameter,
because it determines the amount
of carriers --{\em intrinsic} carriers-- that can be 
thermally excited from the valence band to the conduction 
band. Intrinsic carriers, however, are not very important
at typical device operation temperatures and the conduction
is dominated by {\em extrinsic} carriers, those carriers
that are thermally excited from a dopant level. Hence,
dopants to be such must be very {\em shallow}, meaning that 
the impurity electronic states have to be only a few meV 
from the band edge.

Now, if the band gap broadens as an effect of quantum 
confinement, what happens to dopant levels? In a purely
effective mass picture they will be shifted, like any 
other state, becoming deeper, i.e. the upshift of a 
donor level will be less than of the conduction 
band edge. Clearly, this fact has dramatic consequences 
on the dopant efficiency. 
Namely, a dopant impurity which is known to be very 
shallow in bulk Si, becomes deeper as the diameter 
shrinks down, and it will not eventually be usable
to dope ultra-thin SiNWs. At which diameter does 
this happen? 

From EMT, one can deduce the {\em effective} Bohr-radius 
of the ground state $a_B \approx (\epsilon / m^* ) a_0$,
where $a_0$ is the Bohr-radius of the isolated hydrogen 
atom. This results in about 2.2~nm 
(thus a 4.4~nm diameter) for P. A crude estimate of the 
extension of the wave function is taking twice this 
diameter, thus $\sim$~9~nm. Yet, EMT neglects relaxation 
effect which can be important in very thin SiNWs and the
dopant levels should be calculated directly. 

The trend of the ionization energies vs diameter can be 
qualitatively obtained from DFT calculations~\cite{DurgunPRB07,
LeaoNL08}. As we discussed
in Section~\ref{sub:quantum}, however, the local and
semi-local approximations commonly used for the
exchange-correlation energy severely underestimate 
the band gap and likewise the gap states and the
related ionization energies. While the best suited
approach for correct band gap calculations was the
GW methodology (see Section~\ref{sub:quantum}), it
does not seem a viable solution for defective systems.
Due to the need of simulating isolated impurities,
computational supercells have to be large enough to
allow neglecting the interaction of a defect with its
periodic images. This implies a large number of
atoms which is normally beyond the current computational
capabilities of GW based codes.

Hybrid functionals --where a certain amount of exact 
Hartree-Fock exchange is mixed to conventional LDA/GGA 
functionals-- provide accurate estimations of defect
states in bulk Si~\cite{DeakJPCM05} and have also 
been used to calculate P donors in \iio\ and \iii\ 
SiNWs~\cite{RuraliPRB09}. As expected, these calculations 
yielded ionization energies that are deeper than the
values obtained by DFT~\cite{LeaoNL08}, though
the difference decreases for larger wires. Remarkably,
P behaves as an EMT dopant down to diameters of 1.5~nm,
its wave function highly localized, whereas
it breaks down for wires of 1.0~nm diameter. For such 
small wires the wave function is qualitatively 
different: it significantly interacts with the surface
and cannot be described as quasi-one dimensional 
confined EMT state.

\subsubsection{Dielectric confinement}
\label{sub:dielectric_doping}

The estimation of the ionization energies of dopants in 
nanowires has also been tackled efficiently at the 
tight-binding level~\cite{DiarraPRB07,DiarraJAP08}. We 
have already seen in Section~\ref{sub:quantum} that 
this approach can be complementary to the calculations 
at GW and hybrid-functional level~\cite{NiquetPRB06}. 
Its quantitative reliability could be questioned for 
ultra-thin wires (diameters smaller than 2~nm), because 
this model neglects relaxation effects, 
important for such wires, and relies on a parametrization 
obtained for bulk Si. On the other hand, it is the best 
alternative to deal with larger wires where confinement 
still produces sizeable effects. A remarkable feature of 
this approach is the flexibility with which the screening 
properties of the surrounding dielectric medium can be 
manipulated, allowing to study in detail the so-called 
{\em dielectric confinement}.

The Coulomb potential of an impurity gives rise to a bound 
state in the energy gap. In bulk Si this potential is 
strongly screened ($\epsilon_r=11.3$), the Bohr radius 
is large, and the ionization energies 
amount to a few meV, so that the impurities are ionized 
at room temperature. The screening of a nucleus charge 
$+e$ leaves a total charge $+e/\epsilon$ at the impurity 
site, whereas the remaining charge $+e (1-1/\epsilon)$ is 
repelled at infinity. In a one dimensional system the 
screening properties are different. The charge 
$+e (1-1/\epsilon)$ must be repelled at the surface of 
the nanowire, leading to an extra term in the potential. 

The physics of screening in one dimensional systems is 
straightforwardly incorporated in the tight-binding Hamiltonian 
of \textcite{DiarraPRB07,DiarraJAP08}:
\begin{equation}
H = H_0 + U_{imp} + \sum
\end{equation}
where $H_0$ is the Hamiltonian of the undoped wire, 
$U_{imp}=\pm V(r,r_0)$ is the potential at $r$ of an 
impurity at $r_0$, and $\sum$ is the self-energy potential, 
which accounts for the interaction between the carrier 
and the surface polarization charges induced by its own 
presence.

On the basis of this tight-binding model \textcite{DiarraPRB07} 
showed that the ionization energies of typical 
donors are significantly deeper than in bulk, even for 
large wires ($d>10$~nm) where the effects of quantum 
confinement are weak. This effect is due 
to the interaction between the electron and the surface 
polarization charge $+e (1-1/\epsilon)$. 
These results (i) indicate that dielectric confinement 
can be stronger than quantum confinement and that donor 
levels deepen more than how much the band gap broaden; 
(ii) the dielectric mismatch can be used to vary the 
ionization energies. In particular, a metallic or 
high-permittivity surrounding gate, present in realistic 
applications, is expected to reduce significantly 
the ionization energies. These predictions have been 
recently supported by the experimental results of 
\textcite{BjorkNatNano09}.

\subsubsection{Metallic impurities}
\label{sub:metal_imp}

Although much of the attention has been devoted to
dopants so far, metal impurities are becoming increasingly 
important for the reliability~\cite{HannonNature06,
OhNL08,BaillyNL08,denHertogNL08} and functionalization of SiNWs.
Transition metals have attracted some interest because
of their possible use in designing nanoscale dilute 
magnetic semiconductors~\cite{DurgunPRB08,
GiorgiPRB08,XuJAP08}. Room-temperature ferromagnetism
in SiNWs has indeed been reported~\cite{WuAPL07},
although the annealing conditions to stabilize the
magnetization are very critical. The ferromagnetic
coupling of Mn impurities is confirmed by electronic 
structure calculations. At variance with typical dopants 
these impurities do not segregate to the surface, at least 
in absence of surface DBs~\cite{XuJAP08}, and they favor 
aggregation. This tendency has been reported to be important 
to stabilize the magnetism, because only when the Mn-Mn 
distance is below a certain cutoff ferromagnetic coupling 
is favored over antiferromagnetic coupling~\cite{GiorgiPRB08}.

It is interesting to observe that specific TM impurities 
can drive the nanowire to a half-metallic ground state. 
A half-metal is a system where one spin is metallic, 
whereas the other is insulating~\cite{deGrootPRL83}. 
Such systems have the greatest interest for spintronics 
applications, because naturally all the conduction 
electrons belong to the same spin and the spin 
polarization $P$ is maximum~\footnote{The spin polarization 
is normally defined as $P = (N^{\uparrow} - N^{\downarrow}) / 
(N^{\uparrow} + N^{\downarrow})$, where $N^{\uparrow,
\downarrow}$ are the densities of states at the Fermi 
level.}.
In the framework of a comprehensive analysis of
surface adsorption of TM atoms in SiNWs, \textcite{DurgunPRL07}
discovered that wires decorated with Co and Cr
can be ferromagnetic half-metals. Upon adsorption of a Co or Cr atom
at the surface, the spin degeneracy is lifted and, while 
the bands of majority spin continue being semiconducting, 
two minority spin bands made of hybridized TM-3d and 
Si-3p states cross the Fermi level driving the wire to 
half-metallicity. There is a sizeable charge transfer 
from the TM atom to the wire and the high values of the 
binding energies indicate strong bonds, which is important 
to prevent uncontrolled clustering that would be detrimental 
to the magnetic ordering. The half-metallicity is obtained 
for huge coverages (one impurity for primitive cell) typical 
of dilute magnetic semiconductors. 
As the coverage is reduced, the gap of the minority spins
starts closing and the system is no longer half metallic ($P<1$).
However, the total spin polarization remains very high and
close to its maximum permitted value.

\subsubsection{Formation energy}
\label{sub:form_en}

The energetic cost of creating a defect is given by the 
{\em formation energy}. The formation energy is the main 
computational quantity describing the stability and energetics 
of a defect in a host material and it is essential to determine 
impurity equilibrium concentrations~\cite{ZhangPRL91}, 
solubilities~\cite{VandeWallePRB93} or diffusivities~\cite{FaheyRMP89}.
In bulk systems the formation energy can be calculated 
according to the well established formalism of \textcite{ZhangPRL91}:
\begin{equation}
\Delta E^f = E^D_{tot} - \sum_i n_{i} \mu_{i} + q 
( \varepsilon_v + \mu_e ) , 
\label{eq:ZhangNorthrup}
\end{equation}
where $E^D_{tot}$ is the total energy of the defective 
system, $n_i$ is the number of atoms belonging to species
$i$ with chemical potential $\mu_i$,
$q$ is the net charge of the system, $\varepsilon_v$ 
is the energy of the top of the valence band of the clean 
host and $\mu_e$ is the chemical potential for electrons;
the sum runs over all the species present in the system.
However, the extension of Eq.~\ref{eq:ZhangNorthrup} to 
one dimensional systems presents
some subtleties. In particular, the chemical potential 
of the host species is ill-defined. 

To see why, let us focus on the formation of a Si vacancy and let
us play the movie of the wire growth: Si atoms
start precipitating from the supersaturated Au-Si droplet
and contribute to the nucleation of the wire; once
in a while one Si atom does not {\em fill} the proper 
lattice site and a vacancy is formed. When calculating
the formation energy one has to estimate the contribution 
to the total energy of this atom, the one that left the
vacant site and was incorporated in the wire somewhere else.
In bulk this is easy, because all the lattice sites are 
equivalent and then each atom contributes equally to the
total energy. In a nanowire, on the other hand, this
quantity is not well defined, because it depends where 
the misplaced atom is thought to end up, as the lattice 
sites are non-equivalent.

\textcite{RuraliNL09} have proposed a way to circumvent 
this problem that also deals with the passivating agents,
if present. They showed that rewriting the equations 
for the formation of $N$ defects at a time, being $N$ 
the number of Si atoms in the primitive cell, leads to 
the definition of an {\em effective} chemical potential of the 
wire primitive cell. In this way, whenever the formation 
of a defect involves the addition/removal of an atom of 
the host species, e.g. vacancies, self-interstitials, 
substitutionals, it is transferred to/from the correct 
reservoir --the wire itself-- and only easily computable 
quantities are involved.
 
A further problem arises when dealing with charged defects.
In a periodic boundary condition 
formalism point charges result in an infinite electrostatic 
energy. This inconvenience can be obviated by using a 
compensating jellium background. The errors in the total 
energy are often corrected {\em a posteriori} by means of the 
Madelung energy~\cite{MakovPRB95}, though other methods 
have been proposed. Again, this correction has to be 
properly generalized to be used in nanowires. In particular, 
in solids the Madelung correction is scaled by the value 
of the (isotropic) macroscopic dielectric constant of 
the host material. In nanowires, on the other hand, a 
dielectric tensor $\bar{\bar{\epsilon}}$ will be needed 
for the correct description of the interaction between 
the different instances of the charged defect~\cite{RuraliNL09}.
Notice that the value of the dielectric tensor will depend 
on the ratio between the axial lattice parameter, the 
lateral vacuum buffer and the chosen values of the 
axial~\cite{HamelAPL08} and transverse components of 
the $\bar{\bar{\epsilon}}$ tensor and therefore cannot
be looked up in tables.

\section{Transport properties}
\label{sec:trans}

The study of electron and heat transport is one of the 
most rapidly growing research field in nanowires. 
The reason 
is twofold: on the one hand transport measurements often 
are the most direct and simplest way to test the theoretical 
predictions~\cite{CuiJPCB00,YuJPCB00,SellierPRL06,
BjorkNatNano09,LiAPL03,ChenPRL08}; 
on the other hand, the behavior can be much different from 
bulk Si and can be exploited for enhanced performances in 
applications, whereas other times it can be detrimental. 

\subsection{Electron transport}
\label{sub:trans_elec}

\subsubsection{Surface roughness disorder}

An important cause of the degradation of the electrical 
conductance in SiNW-based devices is the scattering 
occurring at the surface in presence of surface defects 
or surface roughness~\cite{WangAPL05,LuisierAPL07}. 
This is not unexpected, since we have seen that many 
properties are ruled by the large surface-to-volume 
ratio of SiNWs. Considering non-smooth surfaces has indeed
a great importance, as SiNWs exhibiting either 
tapering~\cite{KodambakaPRL06,WangNatNano06,WuPRL08}
or fancier saw-tooth faceting~\cite{RossPRL05}
are often reported.

The effects of surface roughness on electron transport 
have first been addressed by \textcite{SvizhenkoPRB07}. 
They modeled the surface disorder adding with probability 
1/2 one monolayer at each facet of a given unit cell. 
Since the position of these surface bumps is uncorrelated,
they obtained a {\em white-noise} roughness.
The surface roughness originates strong irregularities
in the density of states along the wire axis, which in 
turn causes reflection of carriers and a strong reduction 
of the conductance. 
When these effects sum up in very long wires the disorder 
quickly drives the transport into the Anderson 
localization regime~\cite{AndersonPR58}.

A complementary approach to the description of the
surface roughness disorder has been followed 
by \textcite{PerssonNL08} and \textcite{LherbierPRB08}.
They modeled the roughness as random fluctuations 
$\delta r$ of the wire radius around its average value $r_0$,
through a Lorentzian autocorrelation function, obtaining
a {\em correlated} disorder. They showed that the
backscattering strongly depends on the nanowire 
orientation, the anisotropy coming from the differences 
in the underlying band structure. In particular, electrons 
are less sensitive to surface roughness in \iio\ SiNWs, 
whereas holes are better transmitted in \iii\ 
SiNWs~\cite{PerssonNL08}. Also, as the disorder correlation 
length --roughly the length scale of the diameter 
fluctuations-- increases, the lowest-lying states
of the conduction band get trapped into the largest
sections of the wire~\footnote{A similar mechanism has
been reported recently for selectively strained 
nanowires~\cite{WuNL09}.}. The modified extent of the
electron wave function affect many key quantities 
for transport, such as the mean free path and the 
localization length. Interestingly, the room temperature 
mobility of electrons and holes seems rather insensitive 
to short length scale fluctuations, as well to very long 
length scale fluctuations, a case in which the surface 
experienced by the carriers is locally smooth.

\subsubsection{Single-impurity scattering}

Besides surface disorder, the other main critical source 
of scattering is the presence of impurities. Surface 
scattering has a stronger impact on the transport in 
SiNWs than in bulk Si because of the much larger 
surface-to-volume ratio. The case of impurity scattering 
seems different, since it should solely depend on the 
impurity density and should affect in a similar way bulk 
Si and SiNWs. This is not the case, though. 
So, where is the catch? 

With the reduction of the wire size below 10~nm, 
the impurity cross-sections become of the same order 
of the wire characteristic dimension and can result
in total backscattering. In the semiclassical picture
used to study transport in bulk materials
impurities are point-like centers that scatter randomly
the incoming carriers. This chaotic process slows down
the carrier flow and results in a reduction of the conductance.
The quantum picture in a thin one-dimensional medium is 
slightly different: impurities have to go through to a 
scattering potential that often extends throughout 
most of the wire cross-section and, following with the
semiclassical analogy, the trajectories of the carriers 
are not simply deviated, but they can be entirely backscattered.

{\em Impurity} is a fairly generic denomination when referred
to semiconductors. In fact, it refers to both undesired 
defects, by-products of an imperfect growth,
and to dopants, which are purposely introduced to provide
the material with tailor-made electric features.
Clearly, this case is the most challenging:
dopants increase the carrier density at
device operation temperature, but at the same time
might induce a significant scattering which leads
to a drop in the conductance.
\textcite{Fernandez-SerraNL06} studied the resistance
associated with a substitutional P impurity in the 
wire core, at the wire surface, and with a DB+impurity,
a complex whose importance was discussed in a previous 
work of theirs~\cite{Fernandez-SerraPRL06}. Resonant 
backscattering --a strong reduction of the conductance 
in correspondence to impurity related bound states-- is 
the main signature of substitutional impurities, though 
P in the core or at the surface yield different results. 
On the other hand, DB+impurity complexes are transparent 
to the incoming carriers and the transport is ballistic.

Therefore, donor impurities such as P either segregate 
to the surface where they are likely to form an electrically 
inactive complex with a DB or they stay in the wire core where 
they produce a strong backscattering, particularly at certain 
resonant energies. In both cases the current is reduced. 

\subsubsection{Multiple-impurity scattering}

The calculations of Fern\'{a}ndez-Serra and co-workers 
opened the field of dopant scattering
in SiNWs, but they have two limitations: 
(i)~they study the scattering properties of an individual 
impurity, while in realistic SiNWs the wire resistance
results from multiple scattering events; (ii)~impurities
can be ionized, the typical situation for dopants, and
the proper charge state must be taken into account in the
conductance calculation. 

The first of these two issues
has been tackled by comparing the conductance evaluated 
directly in long wires, with a certain distribution of 
impurities, with the predictions that can be made on the 
basis of single-dopant calculations~\cite{MarkussenPRL07}. 
This is a challenging task, because to make such a comparison 
on equal footing the long wire too has to be treated within 
a first-principle formalism, which involves an 
extraordinary computational load. This difficulty can be 
circumvented thanks to an ingenious method that allows 
constructing the Hamiltonian of the long wire assembling 
building blocks obtained from the single-dopant 
calculations~\cite{MarkussenPRB06}. In this way the 
electronic structure problem has not to be solved directly
in the long wire. 

The surprising result is that the 
properties of long, realistic wires --such as mean free 
path, resistance vs length-- can be {\em entirely} predicted 
from single-impurity conductances. So, the resistance
of a wire with an arbitrary distribution of impurities 
is obtained by classically adding the resistances of
each individual scatterer according to Ohm's law: 
\begin{equation}
\langle R(L,E) \rangle = R_c(E) + \langle R_s(E) \rangle L / l 
\end{equation}
where $\langle R_s(E) \rangle$ is the average resistance 
that can be evaluated from the single-dopant calculations, 
$R_c(E)$ is the contact resistance, $L$ is the length of 
the wire and $l$ the average separation between dopants.
Interestingly, a similar approach has been proposed
also for phonon transport~\cite{SavicPRL08,MarkussenPRB09}.

This method allows easy comparisons of the conductance 
of wires with different distributions of defects. The case 
of P substitutionals, for instance, has been addressed 
by \textcite{MarkussenJCE08}, where a uniform radial 
distribution was compared to a mainly surface distribution, 
in accordance with the previously reported indications of 
P surface segregation~\cite{Fernandez-SerraPRL06}.

\subsubsection{Charged impurity scattering}

Addressing charged impurities poses well-known problems
related to the use of periodic boundary conditions.
Large supercells, out of the current capability of
first-principles methods, are needed to allow the
correct screening of the electrostatic potential of
the impurity. The conductance depends more
critically than other quantities on this incomplete 
screening. 

Such systems have been dealt with within an approximate 
method that combine first-principle methods with finite 
element calculations of the electrostatic 
potential~\cite{RuraliNL08}. The idea is very simple. 
If a charged dopant is approximated with a point charge, 
its electrostatic potential can be obtained in a very 
cheap way with a finite element calculation. Far from the impurity 
this is a reasonable approximation --a P$^+$ impurity 
gives rise to essentially the same Coulomb potential of 
a As$^+$ impurity-- and the agreement with a self-consistent 
electronic structure calculation is indeed very good. 
Close to the impurity, on the other hand, quantum electronic 
structure accounts properly for the different chemical 
nature of different impurities, a task not accomplished 
by a finite element model. 

The part of the potential that converges slowly
with the cell size is the long-range Coulomb potential. 
So, here comes the simple idea: the {\em local} potential
around the impurity is calculated at the first-principle
level in a large, but tractable computational cell. Then
the long-range tails are calculated within a finite element 
model (which it is known to yield the same result than a 
first-principle calculation). In this way one 
can in principle engineer arbitrary boundary conditions,
as the long range part of the potential can be obtained
at a negligible computational cost.
 
This approximated approach allows explicitly addressing
the calculation of the conductance associated to 
impurities with different charge states and comparing
majority 
carriers (electrons in a {\em n}-type wire, holes in 
{\em p}-type wire) vs minority carriers (electrons in 
a {\em p}-type wire, holes in {\em n}-type wire). The 
results are indeed utterly different and it is shown that in 
sufficiently thin wires minority carriers transmission 
is entirely suppressed~\cite{RuraliNL08}.
What happens is that in the case of minority carriers
the dopant constitutes an effective barrier in the
potential landscape. When the energy of the electron
is less than that of the barrier height, it must tunnel 
through the potential and the transmission is therefore 
exponentially suppressed.

\subsection{Heat transport}
\label{sub:trans_heat}

Recently, there has been a lot of excitement around the
thermal conductive properties of SiNWs. Surprisingly, 
this excitement stems from the {\em poor} thermal 
conductance of SiNWs. The reason is that the use of 
SiNWs as materials with enhanced thermoelectric 
properties has been demonstrated independently by two 
groups~\cite{HochbaumNature08,BoukaiNature08}. 

While in some devices one wants to get the heat away as 
efficiently as possible and a high thermal conductance 
is desirable, in thermoelectrics one wants a thermal 
conductance as small as possible~\cite{CahillJAP03}.
It has been suggested that at the nanoscale the thermoelectric
efficiency could be increased with respect to bulk materials, 
since the electrical mobilities are expected to be higher, 
while the surface scattering of the phonons should decrease 
the lattice thermal conductance~\cite{VoNL08}.
Recent experimental results have reported thermal conductivities
of $\sim$~1.6~W~m$^{-1}$~K$^{-1}$, two orders of magnitude lower
than the value for bulk Si (150~W~m$^{-1}$~K$^{-1}$ at room 
temperature).

Heat is transmitted by phonons,
the vibrations of the crystal lattice. Calculating the
phonon modes of a SiNW with a first-principles method is
a demanding task and can be done only for the thinnest 
wires~\cite{PeelaersNL09}. Fortunately, this is not too 
serious an inconvenience and the phonon band structure 
can be calculated with a great level of accuracy within 
simple empirical interatomic potentials. It has been 
shown that phonon dispersions calculated with DFT and 
with the bond-order Tersoff potential~\cite{TersoffPRB89} 
yield thermal conductances in excellent 
agreement~\cite{MarkussenNL08}. It is indeed on the
Stillinger-Weber potential that the first atomistic 
calculations of the thermal conductance of SiNWs were 
based on~\cite{VolzAPL99}.

The decrease of the thermal conductance with the 
reduction of the diameter comes from the interplay of two 
factors: (a)~phonon confinement, i.e. the change in the 
phonon spectra~\cite{AduNL05} 
and (b)~the increase of the inelastic phonon scattering at 
the surface. The dependence of the phonon dispersion 
on the wire size has been studied for \ioo\ wires by 
\textcite{WangAPL07}, who showed that the thermal conductance
decreases as the wire diameter is reduced.
\textcite{PonomarevaNL07b} obtained similar results for
\iii\ wires, although the thermal conductance steeply 
increases again for diameters
below 2~nm~\cite{PonomarevaNL07b}, a direct signature
of phonon confinement. Namely, as the diameter is reduced 
the lowest frequency excited mode is severely affected 
by the confinement; this long wavelength mode dominates 
the low frequency spectrum and, carrying a larger amount 
of energy, determines the enhanced thermal conductance 
at small diameters~\footnote{A similar feature is also present
in the \ioo\ wires of \textcite{WangAPL07}, where for 
temperatures lower than 50~K the thermal conductance of a
1.54~nm wire is larger than the thermal conductance of a
4.61~nm wire. However, it is difficult to be more quantitative
at this respect, due to the scale of the plots used in
\textcite{WangAPL07}.}. 

It should be bear in mind, however, that the effects of
phonon confinement are normally studied in nanowires
with an ideal structure, whereas in the recent reports
of the enhanced thermoelectric figures of SiNWs
surface corrugation seems to play an important role.
Recently, \textcite{DonadioPRL09} showed that the computed thermal 
conductance strongly depends on the surface structure, 
whereas it can be insensitive to variations of the diameter in the
size range investigated ($d<4$~nm). Phonon confinement do
not necessarily lead to low values of the thermal
conductance, which in some cases can even increase as a 
function of size, due to the presence of long wavelength 
phonons with very long mean free paths, like shown by
\textcite{PonomarevaNL07b}.

It is useful to consider these studies together, 
because of their complementary methodological approaches.
\textcite{WangAPL07} calculate the phonon dispersion with 
the usual procedure consisting in displacing each atom 
along $\pm x$, $\pm y$, and $\pm z$ to obtain the dynamical 
matrix by finite differences. Then they calculate the thermal
conductance with non equilibrium Green's functions. In this 
way they rely on the harmonic approximation and thereby 
neglect any phonon-phonon scattering, an approach valid in 
the low to mid temperature limit. Similar calculations 
for thicker wires, $d > 35$~nm~\cite{MingoPRB03,MingoNL03}, 
where phonon confinement effects are unimportant, yielded 
an excellent agreement with experimental results~\cite{LiAPL03}.
\textcite{PonomarevaNL07b} and \textcite{DonadioPRL09}, on 
the other hand, calculate the thermal conductance within 
a molecular dynamics simulation from Fourier's law 
$J_z = -\sigma \partial T / \partial z$, where $J_z$ is 
the heat flow along the wire axis $z$ and $\partial T / 
\partial z$ is the thermal gradient~\cite{SchellingPRB02}.
Within this scheme, one does not calculate explicitly 
the full phonon dispersion and anharmonic effects are 
automatically included in the simulation. The anharmonic 
forces are increasingly important at higher temperatures, 
since the atomic displacements get bigger. This means 
that phonon-phonon scattering becomes more and more 
important at higher temperatures and dominates over the 
effects of including more conducting channels. The drawback 
is that classical molecular dynamics is reliable only above the Debye 
temperature of the material (645~K for Si) where quantum 
effects in the ionic dynamics can be neglected and below 
it the results must be interpreted with some care.

\begin{figure}
\centering
\epsfig{file=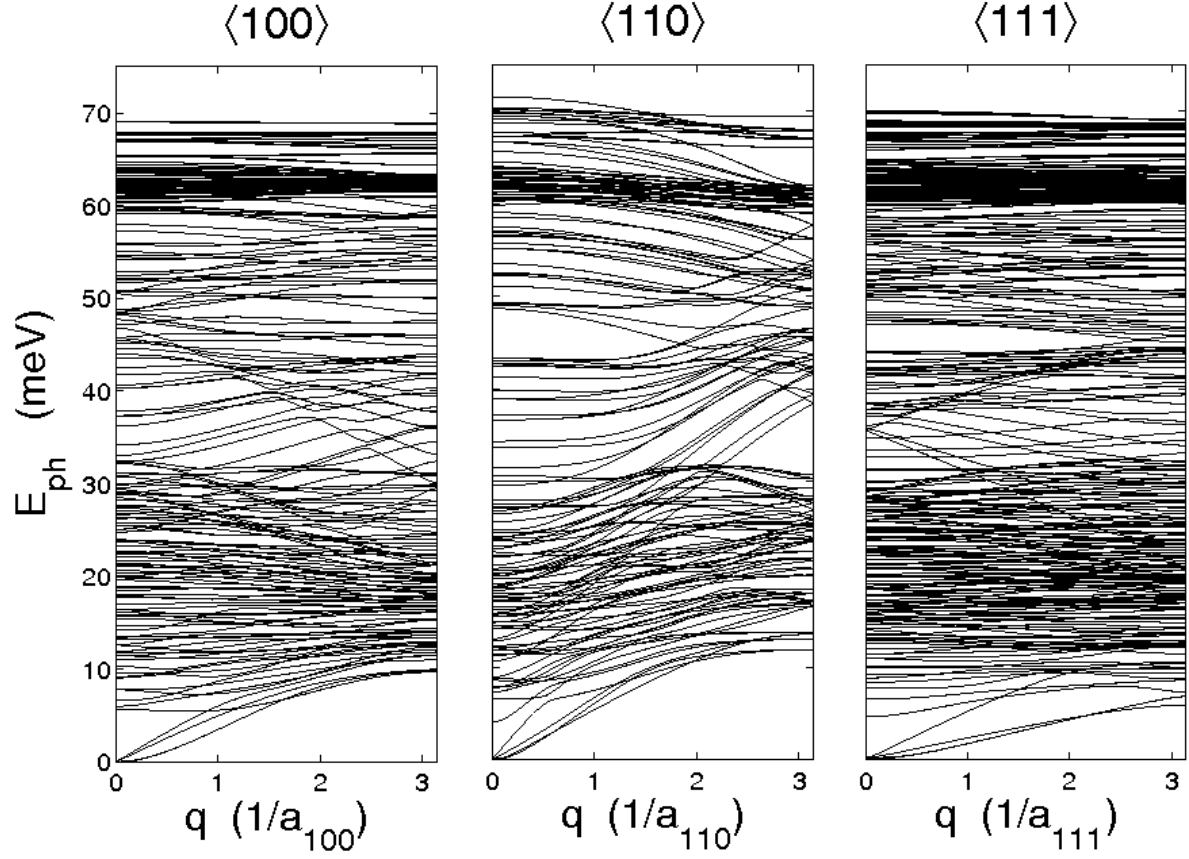, width=1.00\linewidth, clip=, angle=0}
\vskip 0cm
\caption{Phonon band structures calculated within the Tersoff
         bond-order potential of 2~nm diameter SiNWs
         grown along the (a)~\ioo\, (b)~\iio\, and
         (c)~\iii\ axis. The phonon wave vectors,
         $q$, are all in the respective wire directions
         and are shown in units of the reciprocal unit
         cell lengths, with $a_{100}=5.4$, $a_{110}= 3.8$,
         and $a_{111}=9.4$~\AA. From \textcite{MarkussenNL08}.}
\label{fig:MarkussenNL08fig2b}
\end{figure}

It should not come as a surprise that the thermal 
conductance is also anisotropic, like many other
important quantities that we have discussed throughout
the paper.
At low energy the phonon dispersion features four
acoustic branches, one dilatational, one torsional,
and two flexural modes~\cite{ThonhauserPRB04}. The
torsional mode is related to rotational invariance around
the wire axis, and it is similar for all the orientations.
At higher energies, however, one can notice that the
bands in \iio\ SiNWs have a larger slope, i.e.
larger velocities, than the other orientations which
feature mostly flat bands (see Fig.~\ref{fig:MarkussenNL08fig2b}).
Hence, at a given energy there are more bands in the \iio\
wire than in the \ioo\ or \iii\ and consequently one
expects a larger thermal conductance. Heat conductance is
indeed strongly anisotropic. Up to $\approx$~20~K, where
the phonon dispersion is dominated by the acoustic modes,
the thermal conductance is independent on the growth
orientation, but then \iio\ SiNWs stand out, with an up
to twofold increase in their conductance~\cite{MarkussenNL08}.

\section{Conclusions}

In this paper we have reviewed the major advances in the 
theoretical study of the structural, electronic and 
transport properties of silicon nanowires. While the
geometry and the electronic structure of nanowires
are relatively well-understood, many open questions
remain on the possibility of effectively dope ultra-thin
nanowires and on many of the atomic scale mechanisms ruling
electrical current and heat transport. 

Silicon nanowires are rod-like system constructed around
a single-crystalline core. The most important consequence
of their monocrystallinity is that they grow along 
well-defined crystallographic orientations, and at
sufficiently small diameters a strong anisotropy of
most of their properties emerge: the band gap, the
Young's modulus, the electrical conductance or
the specific heat, just to name some, are different
for wires grown along different orientations.
The cross-sections are intimately related with the
growth orientations --given a growth orientation
only certain sets of bounding facets are allowed--,
although their impact on the electronic properties
of the wires seems limited, while other magnitudes
such as the diameter or the surface-to-volume ratio
have a greater influence.
The band gap can be direct, opening the way to 
the use of Si in photonics, and can be tuned by
varying the wire diameter, choosing the growth 
orientation or controlling the surface passivation.
Extrinsic carrier conduction seems to be highly
problematic and whether is feasible or not for
ultra-thin nanowires is not clear yet. The reason
is that dopant efficiency is bedeviled by multiple
factors: surface segregation and clustering with 
consequent neutralization, deepening of the doping
level due to dielectric and quantum confinement. 

Many promising applications have already been demonstrated.
Although the nanowires used in these applications are 
smaller than any device that 
can be fabricated with lithography based techniques, they 
are still larger than those studied theoretically, where
quantum effects leave their clear signature. Whereas it
is clear that nanowires will play an important role in
the next generation of electronic devices, it is difficult
to say if the use of such ultra-thin wires will be practical. 
Many of the properties of these extremely thin nanowires 
pose severe technological challenges, but at the same 
time represent extraordinary opportunities.
Anisotropic band gaps that critically depend on the wire
diameter are apparently incompatible with any standardized 
technological process, to give an example. However, once 
the growth orientation 
and wire thickness can be controlled with great precision
this would open up the possibility of band gap engineering,
which would be extremely attractive for optoelectronics
applications.

Joints efforts in theory and experiments hold the key
to nanowires' future.

\section*{Acknowledgments}
I am deeply indebted to Mads Brandbyge, Xavier 
Cartoix\`{a}, \'{A}d\'{a}m Gali, Nicol\'{a}s Lorente, Anti-Pekka 
Jauho, Troels Markussen, Maurizia Palummo, and Jordi Su\~{n}\'{e}.
I would like to thank all those who granted permission for using 
figures of their original works. 
Financial support by the Ram\'{o}n y Cajal program 
of the Ministerio de Ciencia e Innovaci\'{o}n and funding 
under Contract No.~TEC2006-13731-C02-01 are greatly 
acknowledged. 
\bibliography{../rmp}
\bibliographystyle {apsrmp}

\clearpage

\newpage
\clearpage

\clearpage

\clearpage

\clearpage

\clearpage

\clearpage

\clearpage

\clearpage

\begin{figure}
\centering
\epsfig{file=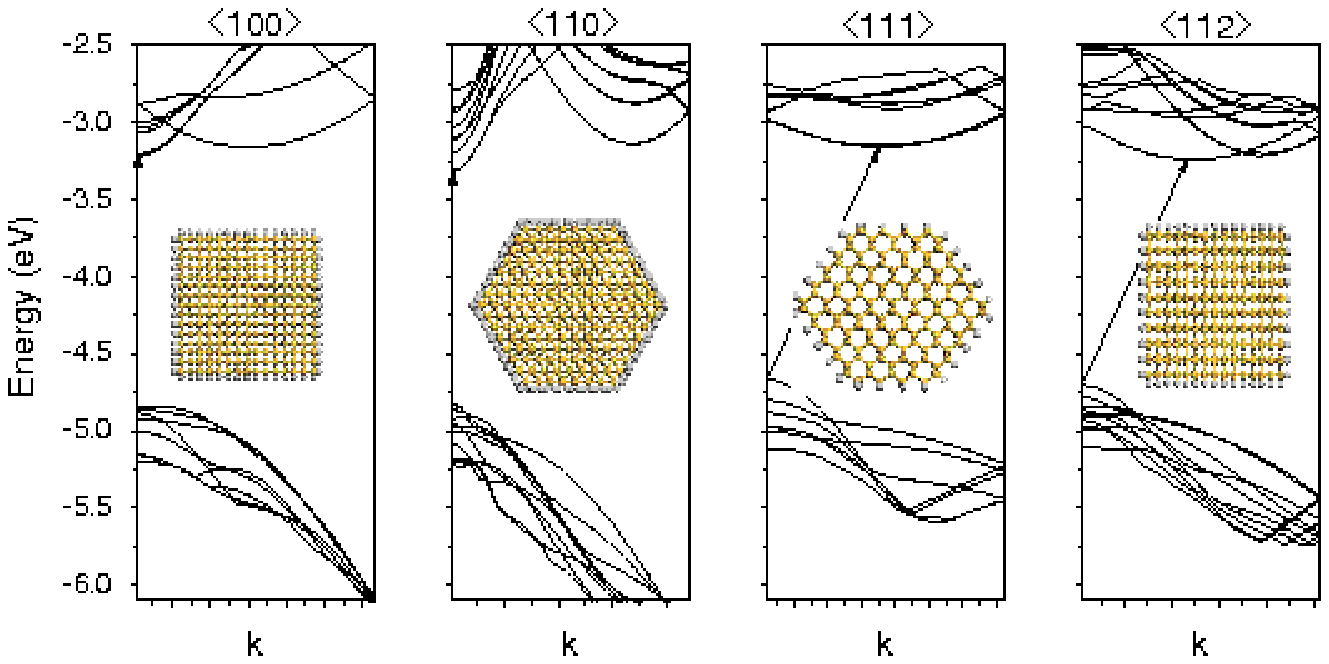, width=1.00\linewidth, clip=, angle=0}
\vskip 0cm
\caption{(Color online) Band structures of \ioo\, \iio\,
         \iii\, and \iiz\ SiNWs with a diameter of
         $\sim 3.0$~nm (cross-sections in the insets).
         The arrows indicate the fundamental band gap
         which is direct for \ioo\ and \iio\ SiNWs and
         indirect for \iii\, and \iiz\ SiNWs. As 
         discussed in the text the band gap of \iii\ 
         SiNWs becomes direct when the diameter is 
         reduced below 2~nm.
         Adapted from \textcite{NgPRB07}. \vskip 250pt}
\label{fig:NgPRB07fig1a-2b}
\end{figure}

\newpage

\end{document}